\documentclass[%
superscriptaddress,
twocolumn,
aps,
pre,
]{revtex4-2}

\usepackage{graphicx}
\usepackage{dcolumn}
\usepackage{bm}
\usepackage[normalem]{ulem}
\usepackage{xcolor} 
\usepackage{hyphenat}
\usepackage{float}
\usepackage{amsmath}
\usepackage{amssymb}
\usepackage{enumitem} 

\usepackage{caption}
\usepackage{subcaption}
\usepackage{ragged2e} 

\usepackage[colorlinks,linkcolor=blue,anchorcolor=pink,filecolor=green,menucolor=yellow,citecolor=blue,urlcolor=blue,bookmarks=false,hypertexnames=true]{hyperref}

\newcommand{\be}{\begin{equation}}
\newcommand{\ee}{\end{equation}}

\definecolor{updateblue}{rgb}{0, 0.0, 0.5}
\newcommand*{\mean}[1]{ \left\langle #1 \right\rangle}
\newcommand{\Juttner}{J\"uttner\;}
\newcommand{\Levy}{L\'evy\;}

\begin{document}

\preprint{APS/123-QED}

\title{
Relativistic \Levy Processes
}

\author{Lucas G.~B. de Souza}
\email{lucas.gabrielsouza@fisica.ufrn.br}
\affiliation{Departamento de {Física} {Teórica} e Experimental, Universidade Federal do Rio Grande do Norte. Natal 59078-970, Brazil}
\affiliation{Institut für Theoretische Physik, Georg-August-Universität Göttingen, Göttingen 37077, Germany}
\author{M.~G.~E. da Luz}
\email{luz@fisica.ufpr.br}
\affiliation{Departamento de {Física} \&
Multidisciplinary Laboratory for Modeling and Analysis
of Data in Complex Systems (MADComplex), Núcleo de 
Modelagem e Computação Científica-Centro Interdisciplinar
de Ciência, Tecnologia e Inovação,
Universidade Federal do {Paraná}. Curitiba 81531-980, Brazil}
\author{E.~P. Raposo}
\email{ernesto.raposo@ufpe.br}
\affiliation{{Laboratório} de {Física} {Teórica} e Computacional, Departamento de {Física}, Universidade Federal de Pernambuco. Recife 50670-901, Brazil}
\author{Evaldo M.~F. Curado}
\email{evaldo@cbpf.br}
\affiliation{Centro Brasileiro de Pesquisas Físicas - CBPF. Rio de Janeiro 22290-180, Brazil}
\affiliation{National Institute of Science and Technology of Complex Systems. Rio de Janeiro 22290-180, Brazil}
\author{G.~M. Viswanathan}
\email{gandhi.viswanathan@gmail.com}
\affiliation{Departamento de {Física} {Teórica} e Experimental, Universidade Federal do Rio Grande do Norte. Natal 59078-970, Brazil}
\affiliation{National Institute of Science and Technology of Complex Systems, Federal University of Rio Grande do Norte. Natal 59078-900, Brazil}
\date{\today}
\begin{abstract}
We study sums of independent and identically distributed random velocities in special relativity. 
We show that the resulting one‑dimensional velocity distributions are not only stable under relativistic velocity addition, but define a genuinely new class of stochastic processes\textendash relativistic \Levy processes.
Given a system, this allows identifying distinct relativistic regimes in terms of the distribution's concavity at the origin and the probability of measuring relativistic velocities.
These features provide a protocol to assess the relevance of stochastic relativistic effects in actual experiments.
As supporting evidence, we find agreement with previous results about heavy-ion diffusion and show that our findings are consistent with the distribution of momentum deviations observed in measurements of antiproton cooling.
\end{abstract}

\maketitle

\section{Introduction}
Constructing a statistical mechanics theory that aligns with the key premises of special relativity (SR) has been a challenge since the formulation of SR in 1905~\cite{einstein1905}.
From a thermodynamic perspective, various attempts have been made to define such a formulation on broader and more general grounds\cite{bergmann1951,hakim1967,AresdeParga2011Sep,farias2017}.
Each scheme tends to make different assumptions for fundamental quantities, establishing the thermodynamic properties accordingly~\cite{farias2017}.
An instructive example relates to the concept of temperature.
Initially, Einstein, Planck and others argued that 
temperature decreases when observed from a moving 
reference frame~\cite{Planck1908Jan}.
Subsequently, from an alternative analysis, H.~Ott concluded exactly the opposite~\cite{Ott1963Feb}.
More recently, a proposal—although not universally favored—is that temperature is a Lorentz scalar~\cite{landsberg1967}, i.e., a real-valued function invariant by the action of the Lorentz transformations.
Further extensions of this idea even maintain that only temperature can be properly defined within each reference frame~\cite{Cubero2007Oct}.
Another approach is to build a thermodynamic equilibrium theory consistent with general relativity and using quantum mechanics as a guiding framework~\cite{rovelli2013}. 
The bottom line is that from a broad  theoretical
viewpoint, these discrepant results are fundamentally
due to the definitions of thermodynamic quantities 
such as heat and work \cite{Dunkel2009Oct,farias2017}.
Significant experiments have been proposed to resolve the issue (see, e.g., Ref.~\cite{Landsberg1980Jul}), but their full execution remains incomplete.

In the realm of statistical physics,
finding {\em bona fide} relativistic distributions is 
not just a mathematical exercise. 
They are crucial for properly interpreting experimental
data. Examples abound in plasma physics~\cite{Dieckmann2006Mar,Sadegzadeh2018Nov}, 
heavy-ion collisions~\cite{Wolschin2004Feb,PHENIXCollaboration2001Dec},
and astrophysics~\cite{Molnar2020Oct,Melrose2021Feb}. 
In cosmology, e.g., the distortion observed in the cosmic
microwave background spectrum~\cite{Uson1988} is explained 
by the Sunyaev-Zeldovich 
effect~\cite{Sunyaev1972Nov,Rephaeli1995}, which
relies on the velocity distribution of electrons
in space~\cite{Itoh1998Jul,Molnar2020Oct}. 
Hence, one needs an accurate expression for this distribution.
The earliest effort in this direction was 
J{\"u}ttner's relativistic generalization of ideal gases~\cite{juttner1911,Juttner1928Jul},
widely employed to characterize several phenomena~\cite{Cercignani,Chacon-Acosta2010Feb,Sadegzadeh2018Nov}. However, it has faced criticism in recent years,
motivating alternative descriptions~\cite{Horwitz1981Dec,Silva2005Nov,Lehmann2006Feb,AresdeParga2011Sep,curado2016,M.F.Curado2022Oct}. 
One of these criticisms concerns the lack of invariance of its functional form over Lorentz transformations~
\cite{M.F.Curado2022Oct}.

\Juttner's original work is based on two central points. 
(1) To assume a flat space of velocities and momenta, giving rise to the volume element $\gamma^{d+2}(v)\, dv$, with $d$ the spatial dimension.
However, this is not a Lorentz-invariant element, i.e., it changes under a Lorentz transformation.
The correct (Lobachevsky-Einstein) relativistic velocity space has a negative curvature~\cite{fock1964}, yielding $\gamma^{d+1}(v) \, dv$, which is Lorentz-invariant. 
(2) The mathematical derivations via the maximization of Boltzmann entropy.
Under such an approach, the relativistic energy is a constant constraint.
This would be reasonable only if the relativistic energy were Lorentz-invariant.

Suppose the Lobachevsky-Einstein velocity space has led to the modified Jüttner (m-Jüttner) distribution~\cite{Dunkel2007Feb,dunkel2007,Aragon-Munoz2018May}.
But (2) above has not been addressed, consequently neglecting the composition law of the dynamical variables (see below).
Therefore, the functional form of Jüttner and m-Jüttner distributions tend to change for each observer, which produces observer-dependent statistics (see Appendix~\ref{appen:invariant_functional_Juttner} for details).
Furthermore, the \Juttner and m-\Juttner distributions disagree with molecular dynamics (MD) simulations of a relativistic gas under certain conditions.
For instance, this kind of simulation has been
performed in Refs.~\cite{curado2016,M.F.Curado2022Oct} 
as well as in Ref.~\cite{Cubero2007Oct}.
Although the latter work validates the \Juttner 
distribution, the former studies show that the \Juttner 
and m-\Juttner distributions deviate from the 
numerical results above a threshold temperature
$\tilde T$.
Explicitly, for the MD velocity 
distribution~\cite{Carroll2019Aug} expressed in terms 
of the rapidity $\sigma(v)$ (see Sec.~\ref{sec2}), 
an unimodal shape is seen for all temperatures $T$. 
Nonetheless, the \Juttner and m-\Juttner
distributions deviate from the MD simulations for 
$T > \tilde T$, when the \Juttner displays bimodal behavior.

As a final relevant remark, relativistic processes are often observed in out-of-equilibrium regimes \cite{Dieckmann2006Mar, Nagaitsev2006Jan}.
Hence, their distributions should differ from those obtained at thermal equilibrium. 
Moreover, several investigations suggest that the stationary states of classical out-of-equilibrium systems can be characterized by stable distributions~\cite{Min1996May,Zanette2003Sep,Barkai2003Nov,Pulvirenti2004Mar,Barkai2004Jun}. 
Thus, constructing stable distributions within the structure of SR would provide valuable insights into a large class of problems.

Motivated by the discussion above, we consider a system composed of a fixed number of $N$ non-interacting particles.
Those exhibit relativistic velocities with sparse collisions, that satisfy relativistic conservation relations and lead to large fluctuations.
%
Nonetheless, contrary to the usual derivations of m-\Juttner (not taking into account the statistical 
properties of the system's degrees of freedom), our construction implicitly\textemdash  through the {\it generalized} central limit theorem (GCLT)~\cite{Nolan2020Sep,levy1925,levy1937}\textemdash assumes an effective noise acting on the relativistic gas. 
Phenomenologically, this could emerge from arbitrary perturbations, e.g., energy injection, regardless of the microscopic model.
The generated fluctuations are captured by the distributions of the particles' dynamical variables.

To describe the system, we derive a relativistic probability distribution by assuming the GCLT in a manner that complies with the relativistic composition of velocities.
This leads directly to a family of one-dimensional 
probability distributions of velocities that are both
statistically stable and have a Lorentz-invariant
functional form.
These distributions allow us to categorize the system's behavior into distinct qualitative velocity regimes by means of simple analyses of their concavity at the origin and the probability of measuring relativistic velocities.
In practice, these regimes are readily identified 
from the values of the distribution parameters.
We further address the implications of our findings on
the statistics of distinct dynamical 
variables, namely, the moments of velocity, 
energy, and momentum.
Additionally, we discuss theoretical and experimental
results supporting the validity of the present 
distributions, presenting good fits for data on 
heavy-ion diffusion~\cite{Wolschin2004Feb} and 
cooling of antiprotons~\cite{Nagaitsev2006Jan}.


\section{Relativistic generalized central limit theorem}
\label{sec2}
The GCLT asserts that a scaled infinite sum of independent
and identically distributed (iid) random variables $w_j$,
namely ${a_\infty \sum_{j} w_j + b_\infty}$
(where ${a_\infty > 0}$, ${b_\infty \in \mathbb{R}}$ and
for the $w_j$s with divergent variance), converges to
the family of one-dimensional L\'evy $\alpha$-stable
distributions in the distribution sense~\cite{Nolan2020Sep}.
The stability index $\alpha \in (0,2)$ governs the asymptotic heavy-tailed power-law behavior of the distribution, with an exponent of $\alpha + 1$.
The boundary value $\alpha = 2$ yields the Gaussian distribution, which also results from the ``standard'' CLT.

In Newtonian physics, we have the Galilean addition rule of velocities ${v = \sum_j v_j}$.
Assuming that $v_j$ represents iid random variables,
the Galilean composition law constitutes the fundamental 
structure of the GCLT, directly enabling the construction
of the L\'evy $\alpha$-stable distribution of velocity $v$, 
with the case $\alpha=2$ yielding the Maxwell-Boltzmann distribution~\cite{maxwellI1860,maxwellII1860}. 
The connection between the CLT and the Maxwell-Boltzmann distribution was established in~\cite{Kinchin1949}, using elements of probability theory, the free particle energy and the independence of the velocity degrees of freedom of the monoatomic gas.
Thus, the GCLT is a general framework for constructing 
stable distributions, provided that a system of multiple components, e.g., particles, is described by physical 
quantities obeying the usual algebraic rule for addition.

\begin{figure*}
\includegraphics[width=\textwidth]{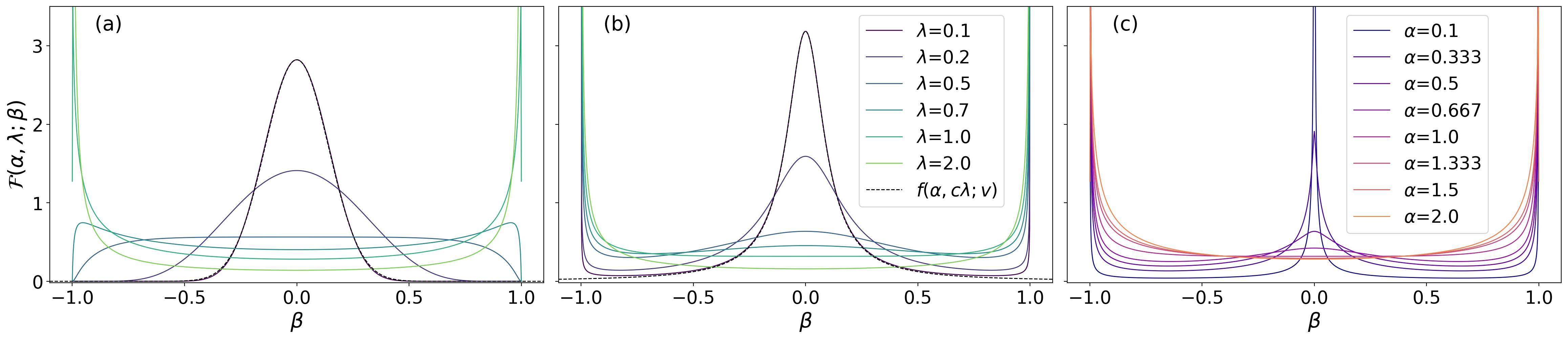}
\caption{\justifying
Relativistic stable distributions of velocities (solid) and their associated non-relativistic stable distributions (dashed). 
The change in modal behavior, characterizing distinct statistical velocity regimes (see main text), is observed as $\alpha$ and $\lambda$ vary.
(a) For $\alpha=2$, at $\lambda=0.1$, the distribution displays a unimodal trend (non-relativistic regime).
As $\lambda$ grows, the distribution transitions to a bimodal shape (relativistic regime), with the threshold being $\lambda > 0.5$.
(b) For $\alpha=1$, even at $\lambda=0.1$, the distribution is trimodal (weak relativistic regime).
(c) For $\lambda=1$, as $\alpha$ increases, the distribution approaches relativistic regime.
The transition to a bimodal shape occurs when $\alpha \geq 1$.
}
\label{fig:relmbcauchy}
\end{figure*}

The \Levy $\alpha$-stable distribution~\cite{levy1925,levy1937,uchaikin1999} reads
$
{f(\alpha,\delta,\nu,\zeta;v) = \int_{-\infty}^{\infty}\frac{dq}{2 \pi} e^{-iqv} \varphi(\alpha,\delta,\nu,\zeta;q)}
$, 
for $\varphi(\alpha,\delta,\nu,\zeta;q)$ being its characteristic function.
The distribution skewness is given by $|\delta| \leq 1$, whereas $\nu$ is the location parameter.
The scale parameter $\zeta \in [0, \infty)$ determines the distribution spread.
Without loss of generality, we focus on the symmetric origin-centered case $\delta=\nu=0$ and, for convenience, we set $\zeta = c \, \lambda$, with $c$ being the speed of light.
Consequently, the traditional, i.e., non-relativistic, stable distribution of velocities reads
\be\label{eq:symstable}
f(\alpha, c \lambda; v) = \int_{-\infty}^{\infty} 
\frac{d q}{2 \pi} 
\, \exp \left(-i \, q \, v - |c \, \lambda \, q|^\alpha\right).
\ee
One of the main features of $f$ is its functional form 
invariance under Galilean transformations.
Suppose that to a reference frame $S$,
the difference $\Delta v_j = v_j - v_{j-1}$ characterizes 
the histogram associated with $f(\alpha, c \lambda; v)$.
But to another reference frame $S'$, of 
relative speed $v'$ with respect to $S$, one gets $v'' = v + v'$ using the Galilean transformation rule.
The new histogram is characterized by 
$\Delta v''_j = v''_j - v''_{j-1} = \Delta v_j$.
This results in the same functional form of $f$, only
with its center shifted to $-v'$. 
It also follows from the GCLT that the distributional
convergence of the infinite sum is not affected by 
adding a scaled random variable proportional to $v'$.
In this sense, the distribution $f$ in Eq.~\eqref{eq:symstable} remains invariant under Galilean transformations.

Now, assume a reference frame $S_1$, with dimensionless
velocity $\beta_1$ with respect to a given inertial 
observer $S$, and a second reference frame $S_2$, with
velocity $\beta_2$ with respect to $S_1$. 
Here $\beta = v/c$. 
Then, in SR the velocity of $S_2$ with respect to 
$S$ is given by the 1D relativistic velocity addition 
defined by~\cite{fock1964}
$\beta = \beta_1 \oplus \beta_2 \equiv (\beta_1 + \beta_2)/(1 + \beta_1 \, \beta_2)$.
This can be extended to $N$ reference frames (see Appendix~\ref{appen:reladdN}) resulting in
\begin{eqnarray}
\label{eq:reladd}
\beta = 
\displaystyle\oplus_{j=1}^{N} \, \beta_j = 
\frac{ \displaystyle \sum_{j \text{ odd}}^N  \, \,
\displaystyle \sum_{n_1 < n_{2} < \cdots < n_j}^{N} 
\beta_{n_1} \dots \beta_{n_j}}{1 +
\displaystyle \sum_{j \text{ even}}^N \, \, 
\sum_{n_1 < n_{2} < \cdots < n_j}^{N}
\beta_{n_1} \dots \beta_{n_j}}.
\end{eqnarray}

\begin{figure*}[ht!]
\includegraphics[width=0.97\textwidth]{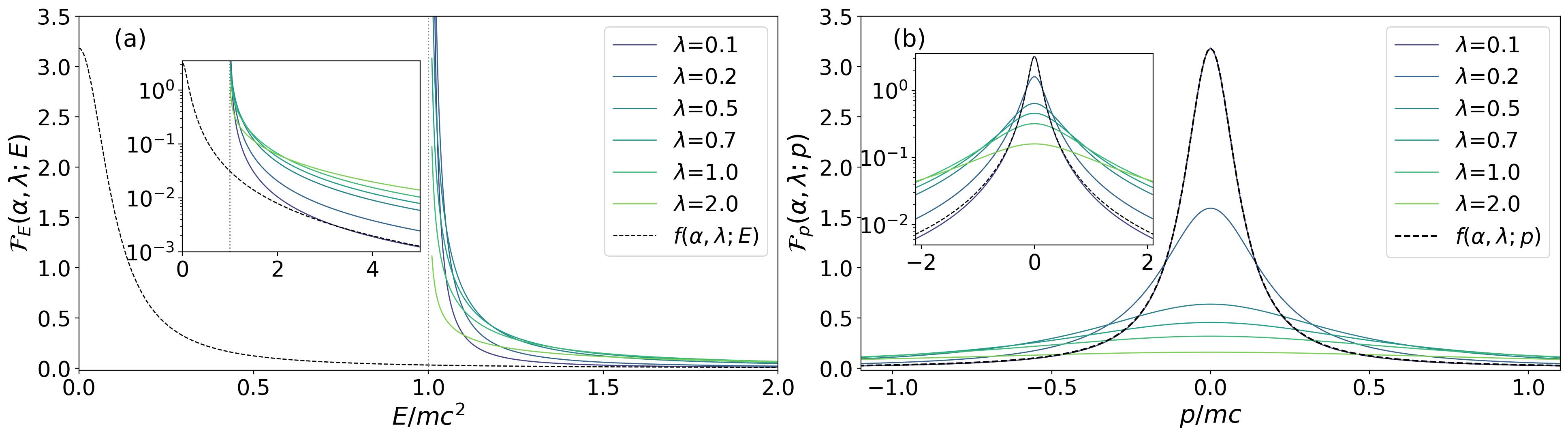}
\caption{\justifying
Relativistic stable distribution of (a)~energy and (b)~momentum 
for $\alpha=1$ (solid) and the corresponding non-relativistic 
cases (dashed).
{The log-linear plots in the insets highlight that these 
distributions exhibit a behavior close to heavy-tail, thus contrasting with 
the distribution of velocities in Fig.~\ref{fig:relmbcauchy}.}
Due to the rest energy, the energy distribution is shifted
by $mc^2$ (dotted line).
}
\label{fig:rel_energy_momentum}
\end{figure*}
In SR one can define the concept of rapidity through
\be\label{eq:rapidity}
\sigma (\beta) = \tanh^{-1} \left( \beta \right) 
= \frac{1}{2} \ln\left( \frac{1+\beta}{1-\beta} \right),
\ee
which essentially represents the angle connecting two
reference frames in the Minkowski
space-time~\cite{Carroll2019Aug}.
Hence, the Lorentz transformation is analogous to 
a rotation in this space-time.
Using this quantity, the relativistic velocity addition in Eq.~\eqref{eq:reladd} can be mapped onto a rapidity addition relation akin to the Galilean addition (see \cite{Rhodes2004Jul} for a similar derivation). 
The relativistic velocity addition ${\beta = \beta_1 \oplus \beta_2}$ can be rewritten as 
$\frac{1+\beta}{1-\beta} = \left(\frac{1+\beta_1}{1-\beta_1}\right)\left(\frac{1+\beta_2}{1-\beta_2}\right).$
From Eq.~\eqref{eq:rapidity}, we obtain
$\sigma(\beta) = \sigma(\beta_1) + \sigma(\beta_2).$
Similarly, the general expression in Eq.~\eqref{eq:reladd} can be rewritten
$$
\frac{1+\beta}{1-\beta} = \left(\frac{1+\beta_1}{1-\beta_1}\right) 
\cdots \left(\frac{1+\beta_N}{1-\beta_N}\right).
$$
Denoting $\sigma(\beta_j)$ by $\sigma_j$, we obtain the addition relation
\be \label{eq:rapaddrel}
\sigma = \displaystyle\sum_{j=1}^N \sigma_j.
\ee

While the Galilean addition law of iid velocity random variables allows calculating their correct statistical distribution, the relativistic composition Eq.~\eqref{eq:reladd} may lead to correlations between the velocities. However, in one dimension the rapidity addition relation is algebraically equivalent to the Galilean rule, making it possible to assume rapidities as iid random variables. 
This makes it possible to preserve the relation between the CLT and the Maxwell-Boltzmann distribution~\cite{Kinchin1949} within the SR framework.
So, by taking the limit $N \to \infty$ and the 
adequate scaling, the $\alpha$-stable distribution 
of $\sigma$ follows straightforwardly from the GCLT,
or
\be\label{eq:stablerapidity}
f(\alpha, \lambda; \sigma) = \int_{-\infty}^{\infty}
 \frac{d k}{2 \pi} \, 
\exp \left(-i \, k \, \sigma - |\lambda \, k|^\alpha\right).
\ee
Moreover, from the relation in Eq.~\eqref{eq:rapidity}, we directly obtain the relativistic $\alpha$-stable 
distribution of velocities $\beta$.
Indeed, recalling the Lorentz factor $\gamma(\beta) = 1/\sqrt{1-\beta^2}$, we find the relativistic $\alpha$-stable distribution $\mathcal{F}$ as
\be \label{eq:rells}
\mathcal{F}(\alpha,\lambda; \beta) = \gamma^2(\beta) \, f(\alpha,\lambda; \sigma(\beta)).
\ee 
Importantly, the GCLT applied to Eq.~\eqref{eq:rapaddrel} returns a distribution whose functional form is invariant
under Lorentz transformations.
Indeed, from $\beta'' = (\beta + \beta')/(1 + \beta \beta')$, 
the rapidity relates as $\sigma(\beta'') = \sigma(\beta) + \sigma(\beta')$, leading to the rapidity histogram 
characterized by $\Delta \sigma_j(\beta'') =\sigma_j(\beta'') - \sigma_{j-1}(\beta'') = \Delta \sigma_j(\beta)$. 
This leads to the same distribution in Eq.~\eqref{eq:rells}, 
centered at rapidity $-\sigma(\beta')$.

To better clarify this key result, in Appendix~\ref{appen:invariant_functional} we explicitly show that under proper construction of a relativistic phase space, the distribution in Eq.~\eqref{eq:rells} can be associated with a Lorentz scalar total probability.
This is only possible if its functional form is invariant under Lorentz transformations~\footnote{
We highlight that, formally, a probability distribution is a representation of a probability measure. 
Hence, the proper representation of ${\cal F}(\alpha, \lambda; \beta)$ is by its measure ${\cal F}(\alpha, \lambda; \beta) \ d\beta$.
Consequently, the total probability associated with this measure, $\int {\cal F}(\alpha, \lambda; \beta) \ d\beta$, is invariant under Lorentz transformations.
}.
Preserving the functional form of Eq.~\eqref{eq:rells} requires adopting Landsberg's perspective~\cite{landsberg1967} of Lorentz scalar temperature.
This translates into the free parameter $\lambda$ scaling with the temperature when ${\alpha=2}$ (see below).

In the Newtonian limit 
${\beta \ll 1}$, ${\gamma(\beta) \approx 1}$ 
and 
${\mathcal{F}(\alpha,\lambda; \beta) \, d \beta \approx f \left( \alpha, c \lambda; v \right) \, d v}$.
So, we retrieve the traditional non-relativistic case.
In the opposite ultra-relativistic situation of $|\beta| \to 1$ (hence, $|\sigma(\beta)| \to \infty$), the asymptotic limit~\cite{Nolan2020Sep} of Eq.~\eqref{eq:rells} with $\alpha<2$ yields
\be\label{eq:rellsasymptotic}
\mathcal{F}(\alpha,\lambda;\beta) \approx \gamma^2(\beta) \, 
\frac{\lambda^\alpha \, \sin \left( \pi \alpha/2 \right) 
\, \Gamma \left(\alpha+1 \right)}{\pi \,
 \sigma^{\alpha+1} (\beta)}.
\ee
While in Eq.~\eqref{eq:symstable} one observes a heavy-tailed behavior as $|v| \to \infty$, Eq.~\eqref{eq:rells} displays peaks near the boundaries $|\beta| \to 1$, as illustrated in Fig.~\ref{fig:relmbcauchy}.

Considering the relativistic expressions for the energy $E=\gamma(\beta) m c^2$ and momentum $p=\gamma(\beta)
\beta  m  c$, we can write 
$\sigma(E) = \cosh^{-1} \left(\frac{E}{m c^2} \right)$
and 
$\sigma(p) = \sinh^{-1} \left(\frac{p}{m c} \right)$,
resulting respectively in
\be \label{eq:rellsenergy}
\mathcal{F}_E (\alpha, \, \lambda; \, E) = 
\frac{f \left(\alpha,\lambda; \sigma(E) \right)}{\sqrt{E^2 - m^2 c^4}}
\ee
and
\be \label{eq:rellsmomentum}
\mathcal{F}_p(\alpha, \, \lambda; \, p) = 
\frac{f \left( \alpha,\lambda; \sigma(p) \right)}{E(p)/c},
\ee
where $E(p)=\gamma(p)mc^2$, with $\gamma(p)=\sqrt{1+(p/mc)^2}$. 
The energy and momentum are not bounded as the velocity, hence the heavy-tail behavior is still present in the
distributions in Eqs.~\eqref{eq:rellsenergy} and \eqref{eq:rellsmomentum}, as shown in Fig.~\ref{fig:rel_energy_momentum}. 
Moreover, due to the rest energy, the relativistic energy distribution is shifted from the origin, as expected.

At this point, a summary of the previous reasoning would be in order.
By construction, the GCLT considering Eq.~\eqref{eq:rapaddrel} leads to a theory consistent 
with SR and whose stability condition is represented by 
the functional form of $\mathcal{F}(\alpha,\lambda; \beta)$.
In other words, an infinite relativistic sum \textemdash as established in Eq.~\eqref{eq:reladd} \textemdash of iid velocity
random variables $\beta$ converges, in the distribution
sense, to Eq.~\eqref{eq:rells}.
The fundamental technical artifice was, 
instead of directly addressing the relativistic addition 
of velocities, to assume the arithmetic addition of 
rapidity, allowing a rigorous and straightforward 
application of the GCLT.
In this way, mapping $\sigma$ back to $\beta$ gives the relativistic velocity \Levy $\alpha$-stable distribution.
This elucidates why $\mathcal{F}(\alpha,\lambda;\beta)$ is given by ordinary $\alpha$-stable distributions (of the rapidity $\sigma(\beta)$) multiplied by the factor $\gamma^2(\beta)$, and is not just a simple change of variables.
From a rigorous standpoint, the validity of the 
relativistic stability condition 
${A_\infty \displaystyle\oplus_{i} \, \beta_{i} + B_\infty}$, 
with ${A_\infty > 0}$ and ${B_\infty \in \mathbb{R}}$, 
follows from the fact that Eq.~\eqref{eq:rapidity} is an isomorphism from the group of relativistic velocities to 
the group of rapidities, constructed with 
Eqs.~\eqref{eq:reladd} and \eqref{eq:rapaddrel},
respectively (see Appendix~\ref{appen:stabcondproof} 
for details).

Finally, we shall comment that the isomorphism property of Eq.~\eqref{eq:rapidity} has been used to derive the m-Jüttner distribution via the maximum entropy principle~\cite{dunkel2007}. 
However, as aforementioned, such construction implicitly neglects the relativistic composition rules~\cite{M.F.Curado2022Oct}.
Further, the relativistic formulation of the CLT has been previously studied~\cite{McKeague2015Apr} using the Kaniadakis $\kappa$-sum~\cite{Kaniadakis2002Nov}. 
Our work extends the results from Ref.~\cite{McKeague2015Apr} to a functional form of stable relativistic distributions without introducing arbitrary, i.e., formal rather than physical, constraints.

\subsection{Relevant particular cases}
\begin{figure}
\includegraphics[width=0.47\textwidth]{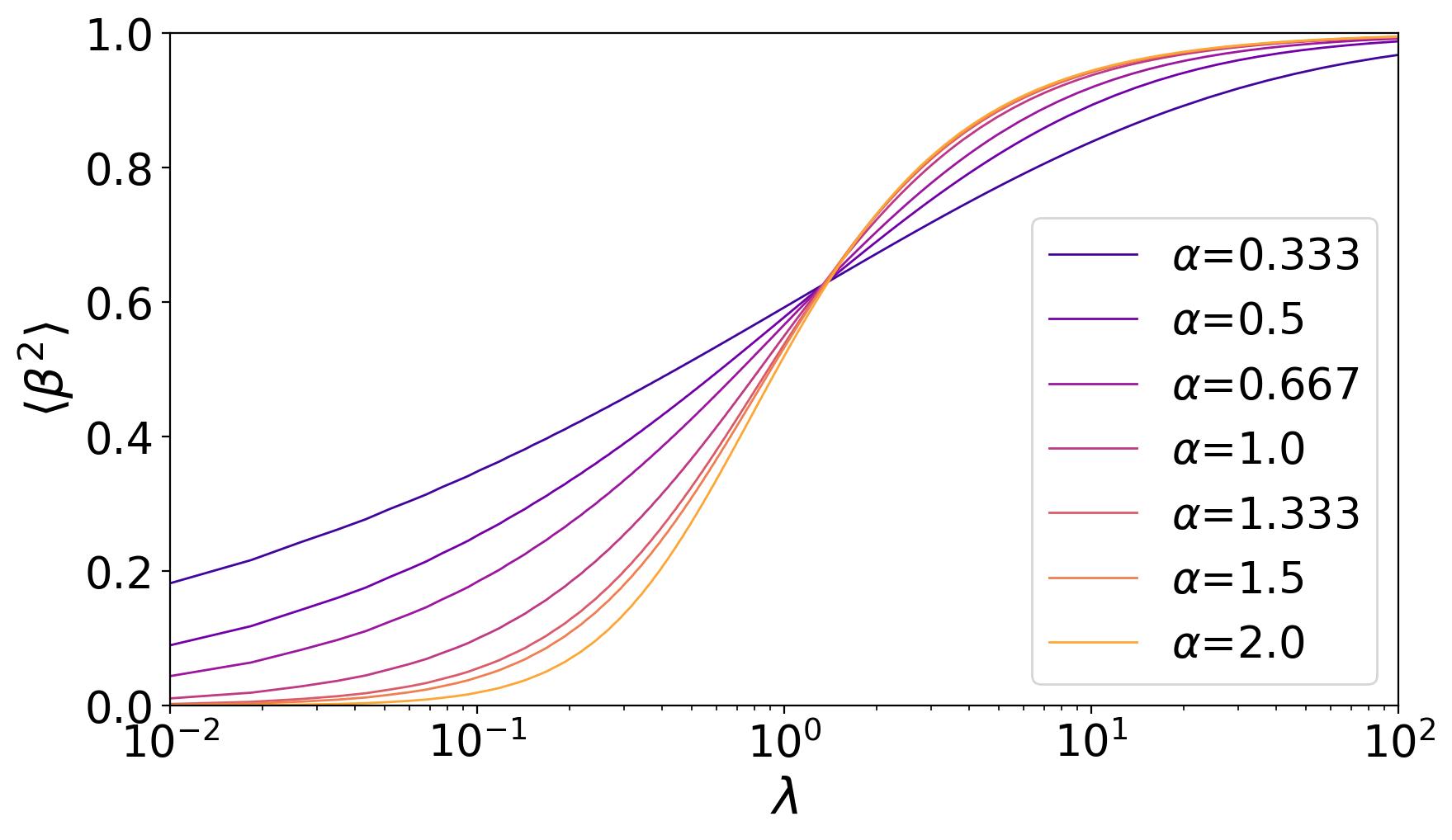}
\caption{\justifying
Variance of relativistic velocity, $\langle \beta^2 \rangle$, as a function of the scale parameter $\lambda$ for several values of $\alpha$.
Because the velocity is bounded by $c$, the variance converges as $\lambda \to \infty$, irrespective of $\alpha$.
}
\label{fig:meanv2rellsalphas}
\end{figure}

A key property of \Levy $\alpha$-stable distributions is that their variance diverges for $\alpha<2$, and their mean diverges for $\alpha<1$.
Nonetheless, since in SR the speed of light bounds $v$, the $q$th moment $\mean{ v^q }$ is bounded by $c^q$.
Specifically, the $q$th moment ${\mean{\beta^q} = \int_{-1}^{1} \beta^q \mathcal{F}(\alpha,\lambda;\beta) d\beta}$ vanishes for odd values of $q$ due to the distribution’s symmetry and is bounded by $1$ otherwise.
As depicted in Fig.~\ref{fig:meanv2rellsalphas}, $\mean{\beta^2}$ converges to $1$ as $\lambda$ increases. 
This behavior holds for any $\alpha$.
In contrast, the moments of Eqs.~\eqref{eq:rellsenergy} and \eqref{eq:rellsmomentum} are finite only when $\alpha = 2$, as a consequence of heavy-tailed behavior.

Note that when $\alpha=1$ in Eq.~\eqref{eq:rells}, we get the relativistic Cauchy distribution of velocities
$$
\mathcal{F}(1,\lambda;\beta) = \gamma^2(\beta) \frac{\lambda/\pi}{(\sigma^2(\beta)+\lambda^2)}.
$$ 
In this case, the variance is bounded and given by
$
{\mean{ \beta^2 } = 1 - \frac{2}{\pi^2} \psi^{(1)} \left(\frac{\lambda}{\pi} + \frac{1}{2} \right)},
$
where 
$
\psi^{(1)}(z) = \frac{ \mathrm{d}^{2} }{ \mathrm{d}z^{2} } \ln(\Gamma (z))
$ 
is the first-order polygamma function~\cite{erdelyi1953} (see Appendix~\ref{appen:meanv2} for derivation).

For $k_B$ the Boltzmann constant and $T$ the temperature, setting $\alpha=2$ and $\lambda^2 = k_B T / 2mc^2$ in Eq.~\eqref{eq:symstable} recovers the Maxwell-Boltzmann distribution, whereas Eq.~\eqref{eq:rells} yields its relativistic version
$$\mathcal{F} (2, \sqrt{T}; \beta) = \sqrt{\frac{mc^2}{2\pi k_B T}} \, 
\gamma^2(\beta) \exp\left(-\frac{mc^2\sigma^2(\beta)}{2k_B T}\right).$$ 
This result was derived in~\cite{curado2016,M.F.Curado2022Oct} as an alternative to the m-\Juttner distribution.

We can also define the $q$th moments of the energy and momentum, respectively, as
$
\mean{E^q} = \int_{-1}^{1} (\gamma(\beta) \, m \, c^2)^q 
\, \mathcal{F}(\alpha,\lambda;\beta) \, d \beta
$ 
and 
$
\mean{p^q} =\int_{-1}^{1} (\gamma(\beta) \, \beta \, m \, c)^q 
\, \mathcal{F}(\alpha,\lambda;\beta) \, d \beta
$.
For $\alpha=2$, we obtain 
\be
\mean{E} = m c^2 \exp \left(\frac{k_B T}{2 m c^2} \right),
\ee
\be
\mean{E^2} = \frac{m^2 c^4}{2} \left(
1 + \exp\left(\frac{2 k_B T}{mc^2}\right) \right),
\ee
and  
\be
\mean{p^2} = \frac{m^2 c^2}{2} \left(\exp \left(\frac{2 k_B T}{ m c^2}
\right) -1 \right).
\ee
These expressions yield the expected Lorentz-invariant relation 
${\mean{E^2} = \mean{p^2} c^2 + m^2 c^4}$.
Furthermore, for small $T$, it reads
${\mean{p^2}=m^2 \mean{v^2}}$ and
${\mean{E} - mc^2 = \frac{m\mean{v^2}}{2}}$, which corresponds to the Newtonian limit.

\section{Relativistic effects quantifiers}
\subsection{Concavity at the origin}
\begin{figure*}
\begin{minipage}{0.75\textwidth}
\includegraphics[width=0.49\textwidth]{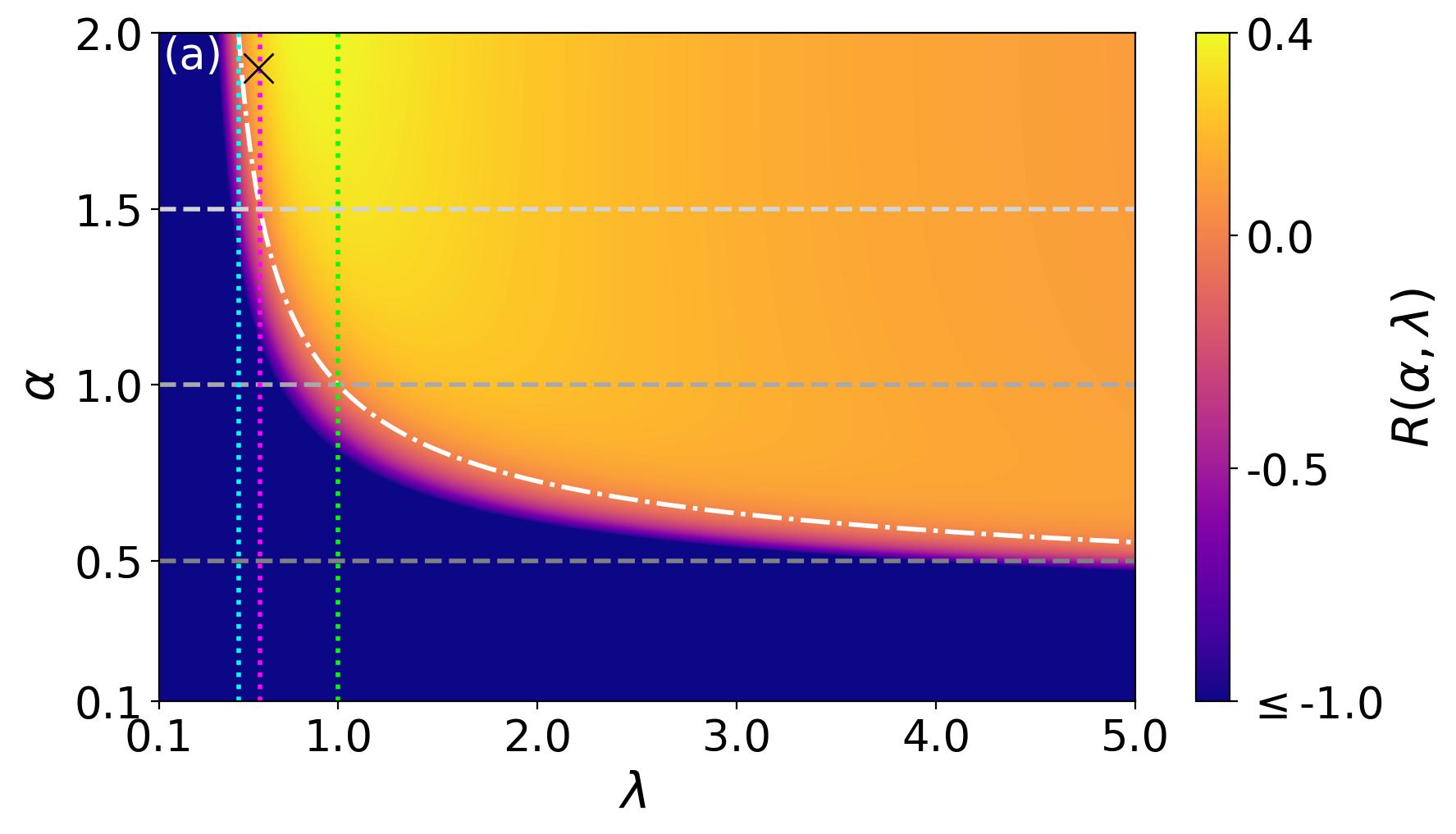}
\includegraphics[width=0.48\textwidth]{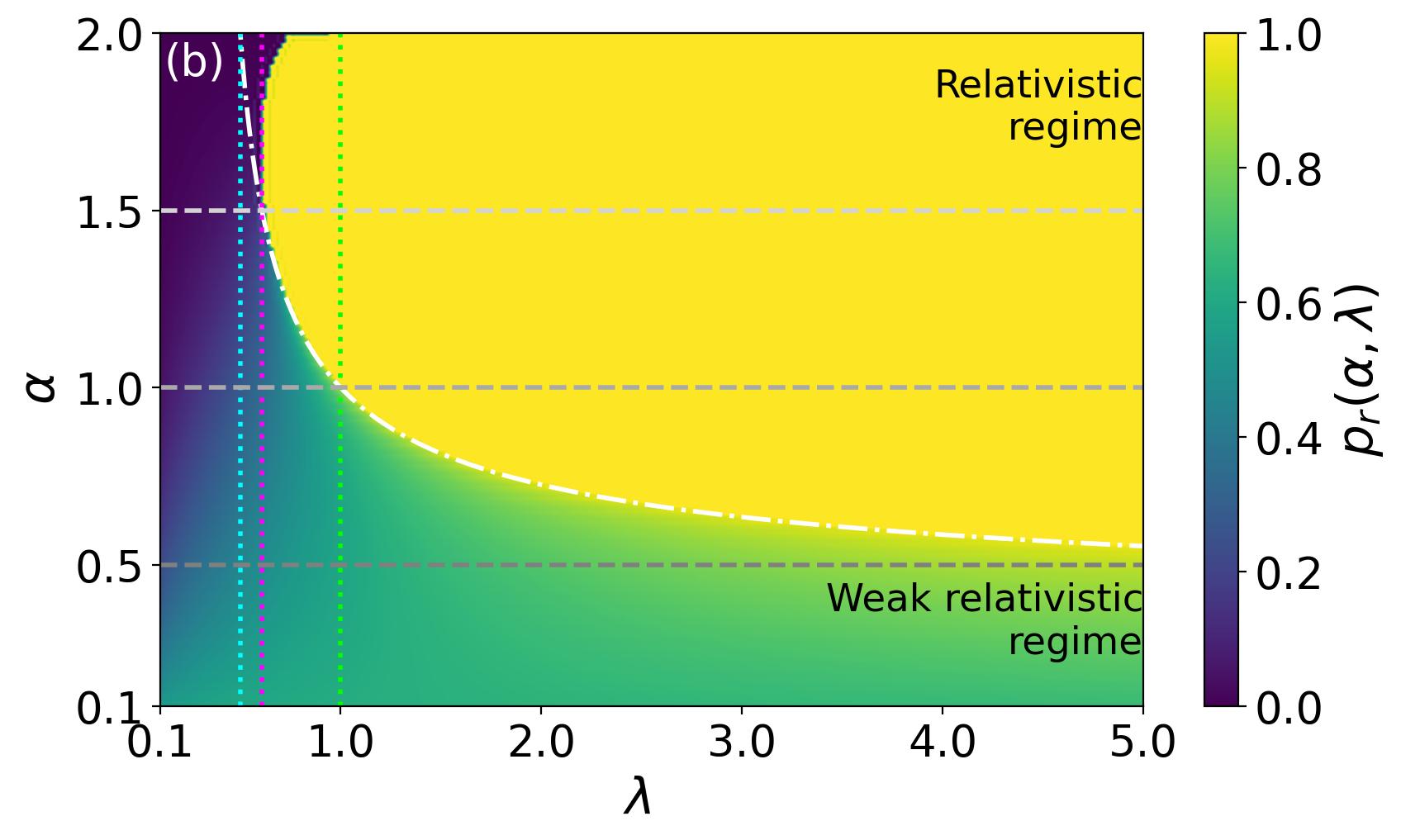}
\end{minipage}
\begin{minipage}{0.24\textwidth}
\begin{subfigure}{\textwidth}
\includegraphics[width=\textwidth]{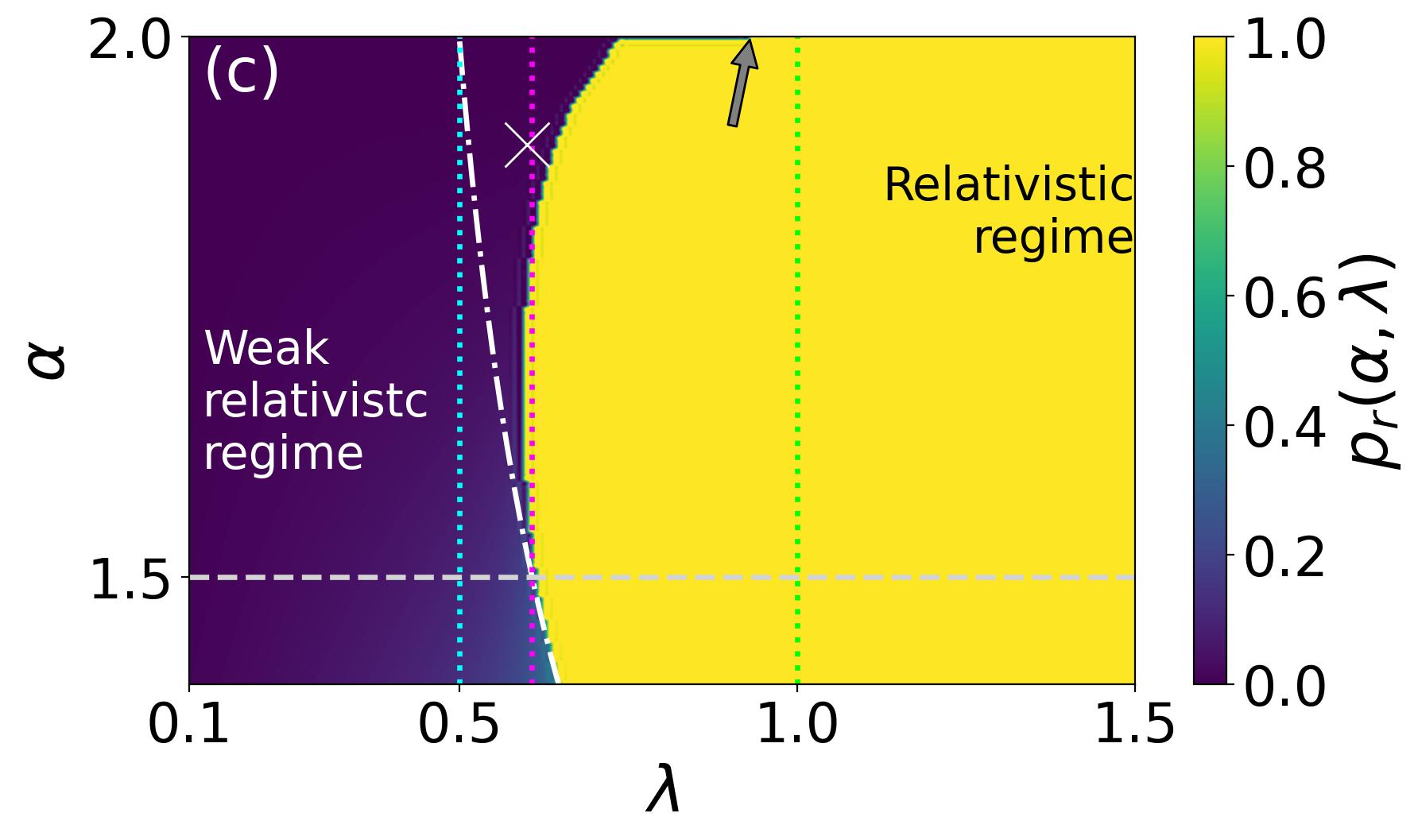}    
\end{subfigure}
\vfill
\begin{subfigure}{\textwidth}
\includegraphics[width=\textwidth]{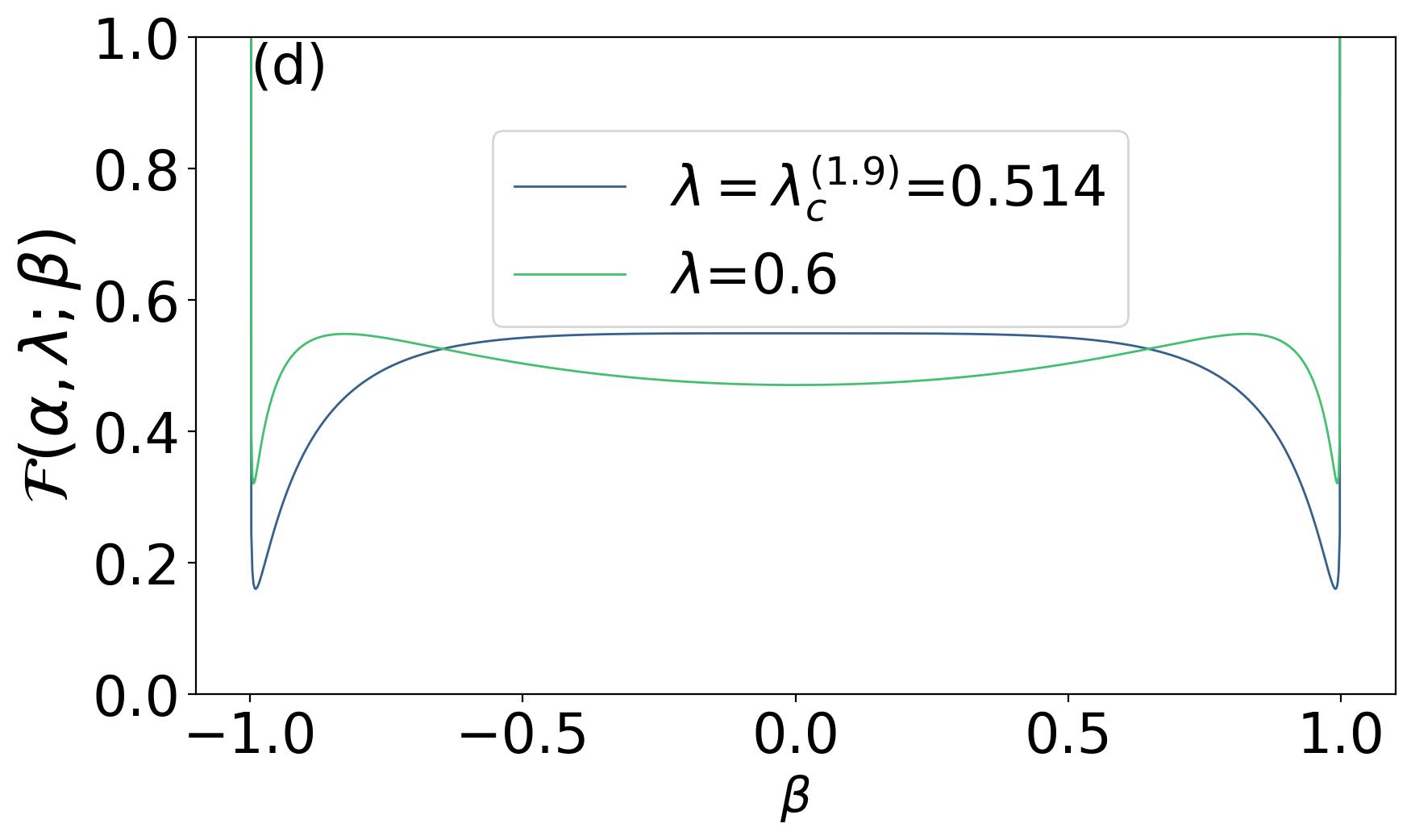}  
\end{subfigure}
\end{minipage}
\caption{\justifying
(a)-(c) Density plots of the quantifiers $R$ and $p_{r}$ (the latter is reliable for any value of $\alpha$).
(d) Distributions for the parameters ($\lambda=0.6$, $\alpha=1.9$) and ($\lambda = \lambda_c^{(1.9)} = 0.514$, $\alpha=1.9$) \textemdash indicated by $\times$ in (a) and (c) \textemdash displaying a special quadrimodal shape for the former, a trend that cannot be predicted solely by the concavity at $\beta=0$, given by $R$.
The (white) dotted-dashed curves represent the set of points $(\lambda_c^{(\alpha)},\alpha)$. In (b), for $\alpha < 2$, the border ${\mathfrak B}$ between the yellow and the other colored regions marks the transition between the {relativistic ($p_r=1$) and weak relativistic ($p_r < 1$)} regimes.
The vertical and horizontal lines indicate special parameter values and are guides to the eye.
A blow-up in the region of $\alpha > 1.4$ displayed in (c) evidences a mismatch between $R$ and $p_{r}$ in characterizing the transition from weak relativistic to relativistic regime when $\alpha>1.5$.
While $R$ predicts a continuous transition from {weak relativistic to relativistic regime} for any $\alpha$, $p_{r}$ predicts a continuous (discrete) transition for $\alpha<1.5$ ($\alpha \geq 1.5$).
Also, for $\alpha=2$, $\lambda_c^{(2)} = 0.5$, but the transition takes place for $\lambda \approx 0.9$, as indicated by an arrow in (c).
}
\label{fig:relvelprobcontribution}
\end{figure*}
In Fig.~\ref{fig:relmbcauchy}(a), we observe that ${\cal F}(\alpha,\lambda;\beta)$ transitions from high to low-temperature regime as $\lambda$ decreases, in which the velocities are insufficiently high to display relativistic effects.
Consequently, as it should be, Eq.~\eqref{eq:rells} with 
$\alpha=2$ recovers the usual Maxwell-Boltzmann distribution.
Visually, this is characterized by the shift from a bimodal to an unimodal shape of the distribution \textemdash notice the disappearance of peaks at $|\beta| \to 1$ and the emergence of a peak at $\beta=0$.
In Fig.~\ref{fig:relmbcauchy}(b), a different scenario occurs for $\alpha=1$, where a transition from a bimodal to a trimodal behavior emerges as $\lambda$ decreases. 
This behavior is observed for other $\alpha<2$ values as well.
It can be checked from the analytic form of Eq.~\eqref{eq:rells} attained when $\alpha$ is rational (see Appendix~\ref{appen:alpharational}).
These findings show that for $\alpha<2$, the non-relativistic regime cannot be reached.
As seen in Fig.~\ref{fig:relmbcauchy}(b), even at $\lambda=0.1$ the peaks at $|\beta| \to 1$ tend to persist, although with an onset closer to $|\beta| = 1$.
Figure~\ref{fig:relmbcauchy}(c) shows that the latter behavior is intrinsic to the relativistic stable
distribution of velocities when $\alpha<2$.
Specifically, with $\lambda=1$, as $\alpha$ increases the distribution changes from a trimodal to a bimodal trend.
This phenomenon is a consequence of the peaks occurring 
at $|\beta| \to 1$ when $\alpha<2$, which disappear only at $\lambda=0$ (see Eq.~\eqref{eq:rellsasymptotic}).
Hence, in all scenarios far from the Gaussian distribution ($\alpha=2$), the system will typically exhibit a bimodal or trimodal distribution depending on $\lambda$, indicating the statistical presence\textemdash even if small\textemdash of relativistic velocities and so relativistic effects. 

One can quantify these traits directly from  Eq.~\eqref{eq:rells}. 
Note that the concavity $R(\alpha,\lambda) \equiv \mathcal{F}''(\alpha,\lambda;\beta) \mid_{\beta=0}$ is a quantifier measuring the relativistic effects once it specifies the velocity regimes of Eq.~\eqref{eq:rells}.
For $\alpha < 2$, if the distribution in Eq.~\eqref{eq:rells} is trimodal, it indicates a large probability of measuring non-relativistic velocities.
In addition, the peaks at $|\beta| \to 1$ suggest a substantial probability of also measuring relativistic velocities.
In the bimodal behavior, it is suggested that there is a high probability of measuring mostly relativistic velocities.
So, $\mathcal{F}(\alpha,\lambda;\beta)$ is in the weak (normal) relativistic regime when the trimodal (bimodal) behavior takes place.
Consequently, we can have a classification\textemdash only when $\alpha < 2$\textemdash based on the sign of the concavity function
$R(\alpha,\lambda)$: 
$R(\alpha,\lambda) < 0$ 
($R(\alpha,\lambda) \geq 0$)
usually characterizes the weak (normal) relativistic regime.
Using Eq.~\eqref{eq:rells}, we find
\be\label{eq:rellsconcavity}
R(\alpha,\lambda) = \frac{2}{\pi \, c^2 \, \lambda}
\left( \Gamma(1+\alpha^{-1})
- \frac{\Gamma(1 + 3 \, \alpha^{-1})}{6 \, \lambda^2} \right).
\ee
The behavior of $R$ is depicted in Fig.~\ref{fig:relvelprobcontribution}(a).
From $R$ we can define a critical value 
$\lambda_{c}^{(\alpha)}$ of the scale parameter, at which $R(\alpha,\lambda_{c}^{(\alpha)}) = 0$, or
$$
\lambda_{c}^{(\alpha)} = \sqrt{ \frac{ \Gamma(1 + 3 \, \alpha^{-1}) } { 6 \, \Gamma(1+\alpha^{-1}) } }.
$$
The curve generated by the points $(\lambda_{c}^{(\alpha)},\alpha)$ is shown in Fig.~\ref{fig:relvelprobcontribution}(a)--(c).
Note that for $\alpha=2$, $\lambda_{c}^{(2)}=0.5$, and for $\alpha=1$, $\lambda_{c}^{(1)}=1$ (see dotted lines in Fig.~\ref{fig:relvelprobcontribution}), both are easily identified in Fig.~\ref{fig:relmbcauchy}.
From Fig.~\ref{fig:relvelprobcontribution}(a) we see that
$\lambda=\lambda_{c}^{(2)}=0.5$ is the minimum possible value, with $\lambda_{c}^{(\alpha)}$ increasing with $\alpha$ decreasing.
Thus, for $\lambda \leq 0.5$ only weak relativistic ($\alpha < 2$) and non-relativistic ($\alpha = 2$) regimes are possible.

However, some care is necessary in using this quantifier exclusively.
For instance, it predicts the possibility of relativistic regimes for any $\alpha$ and $\lambda > 0.5$, but rather the correct condition is $\lambda > \lambda_c^{(1.5)} \approx 0.608$ (dotted line in Fig.~\ref{fig:relvelprobcontribution}), as we are going to see next.
Also, as already mentioned, the cases of $\alpha$ close or equal to $2$ constitute exceptions, with a direct transition from non-relativistic (unimodal) to relativistic (bimodal) regime taking place for $\alpha=2$~\cite{M.F.Curado2022Oct} (observe the transition from $\lambda=0.2$ to $\lambda=1$ in Fig.~\ref{fig:relmbcauchy}(a)).
This is not well-evidenced by Eq.~\eqref{eq:rellsconcavity}.
In particular, we have that for $\alpha=2$ only when $\lambda>0.9$ is the relativistic regime achieved (refer to Fig.~\ref{fig:relvelprobcontribution}(c) and the discussion below), showing that $R$ and $\lambda_{c}^{(\alpha)}$ are not suitable to study the $\alpha=2$ case.
Finally, when we specifically consider $0.5 < \lambda < 0.7$ and $\alpha > 1.5$, some extra aspects of the distributions are not fully captured by $R(\alpha,\lambda)$. 
For example, a quadrimodal behavior can also emerge, 
as shown in Fig.~\ref{fig:relvelprobcontribution}(d), 
making it difficult to characterize the actual distribution's regime.
Hence, in addition to $R$ and $\lambda_c^{(\alpha)}$, below we propose another quantifier to typify the relativistic \Levy stable distributions.


\subsection{Probability of measuring relativistic velocities}
The distribution in Eq.~\eqref{eq:symstable} displays
a decreasing trait with $v$.
Yet in SR, for most combinations of $\alpha$ and $\lambda$,
corresponding Eq.~\eqref{eq:rells} tends to increase
as $|\beta|$ approaches $1$ (see Fig.~\ref{fig:relmbcauchy} 
and Sec. \ref{secquatro}).
Therefore, except for the unimodal behavior (non-relativistic), there must be a particular value of $|\beta|$, $0 < \beta_{\alpha,\lambda} < 1$, for which the distribution exhibits a minimum.
This is clearly illustrated in the asymptotic expression in Eq.~\eqref{eq:rellsasymptotic}.
The terms $1/\sigma(\beta)^{\alpha+1}$ and $\gamma^2(\beta)$, respectively, steadily decrease and increase with $|\beta|$, leading to points of minimum $\pm \beta_{\alpha,\lambda}$ 
at which $\gamma^2(\beta)$ starts to dominate, thus marking the onset of relativistic effects.
Then, we can introduce the probability of measuring 
relativistic velocities as 
\be \label{eq:velprob}
p_{r}(\alpha,\lambda) = 1 - \int_{-\beta_{\alpha,\lambda}}^{\beta_{\alpha,\lambda}}
\mathcal{F}(\alpha,\lambda;\beta) \, d\beta.
\ee 
Notice that for the Newtonian limit, 
$\mathcal{F}(\alpha,\lambda;\beta) \, d\beta \approx f(\alpha,c\lambda;v) \, dv$ 
and 
$\beta_{\alpha,\lambda} \rightarrow 1$, 
so that $p_{r} = 0$.

The variation of $p_{r}$ with $\alpha$ and $\lambda$ is displayed in Fig.~\ref{fig:relvelprobcontribution}(b), with a detailed view of the interval $1.4 \leq \alpha < 2.0$ presented in Fig.~\ref{fig:relvelprobcontribution}(c).
For $\alpha < 2$ we clearly see a border ${\mathfrak B}$ between the weak relativistic ($p_r < 1$) and relativistic ($p_r = 1$) regions. 
We also show the collection of points 
($\lambda_c^{(\alpha)}, \alpha)$, generating the dot-dashed curve.
Importantly, the $(\lambda_c^{(\alpha)}, \alpha)$ curve 
coincides with ${\mathfrak B}$ for $\alpha \leq 1.5$. 
Nonetheless, for $\alpha > 1.5$, it deviates to the left of ${\mathfrak B}$, i.e., to the relativistic region (see Fig.~\ref{fig:relvelprobcontribution}(c)), 
illustrating the limitations of the quantifier $\lambda_c^{(\alpha)}$ depending on the parameter ranges.

For the region $\lambda < \lambda_c^{(1.5)} \approx 0.608$
and $\alpha < 2$, $\mathcal{F}(\alpha,\lambda;\beta)$
always falls into the weak relativistic regime.
But as $\alpha$ increases, particularly for $\alpha>1$,
$p_{r}$ decreases, implying the weak relativistic regime close to the non-relativistic case of Eq.~\eqref{eq:symstable}.
Further, for $0.1 \leq \alpha \leq 0.5$, the {relativistic 
regime} cannot be attained in the considered range of
$\lambda \leq 5$ given that for these $\alpha$ values, $\lambda_c^{(\alpha)}$ lies between $\lambda_{c}^{(0.5)}=7.75$
and $\lambda_{c}^{(0.1)}= 3.5 \times 10^{12}$.
This is visually seen from the trimodal shape of 
${\mathcal F}(\lambda=0.1)$ in 
Fig.~\ref{fig:relmbcauchy}(b), which is quite similar to $f$ for almost all values of $\beta$.
Thus, $p_{r}$ essentially represents the area of ${\mathcal F}$ for $|\beta|$ close to 1.
In contrast, for $\lambda \geq 0.608$ we have that for most of the $(\lambda,\alpha)$ space, $p_{r}$ increases with $\alpha$.
The exception is in the region highlighted in Fig.~\ref{fig:relvelprobcontribution}(c), just when ${\mathfrak B}$ does not coincide with the $(\lambda_c^{(\alpha)}, \alpha)$ curve.
We note that similar analyses~\cite{Mendoza2012Aug} have been conducted for the Jüttner distribution, employing a different method and using the temperature, rather than $\alpha$ and $\lambda$, as the control parameter.

As previously observed, $R$ also may serve as a quantifier of the different distribution regimes, provided the ranges for $\alpha$ and $\lambda$ are properly considered. This becomes clear by comparing Fig.~\ref{fig:relvelprobcontribution}(a) and (b).
Indeed, for $\alpha \leq 1.5$ the overall patterns in both figures are akin. 
Moreover, the sign changes of $R$ correctly match the 
weak to the normal relativistic regimes transition of $p_r$.
For instance, consider $\alpha=1$ in Fig.~\ref{fig:relvelprobcontribution}(a) and (b), 
indicated by dashed lines.
Following these lines from $\lambda=0.1$ to the critical value $\lambda = \lambda_{c}^{(1)} = 1$, we observe that $R$ and $p_{r}$ steadily increase until $0$ and $1$, respectively. For $\lambda > 1$, we have $R > 0$ and $p_r = 1$, thus both indicating a transition to the {relativistic regime}.
Nevertheless, the range $\alpha > 1.5$ should be considered more carefully.
Now, $\beta_{\alpha,\lambda}$ and $p_{r}$ still appropriately characterize the different regimes, but $R$ \textemdash determining the number of modes of the distributions \textemdash by itself is no longer enough to typify the distribution's global behavior.
This manifests in the differences between Fig.~\ref{fig:relvelprobcontribution}(a) and (b), emphasized in Fig.~\ref{fig:relvelprobcontribution}(c).
As an example, Fig.~\ref{fig:relvelprobcontribution}(d) 
shows the distribution for $\alpha = 1.9$, and 
$\lambda=0.6 \text{ and } 0.514$. 
Note that for $\lambda=0.6$, $p_{r}$ approaches zero. 
However, $R$ predicts it should be one since $\lambda=0.6$ is above the critical value $\lambda_c^{(1.9)}=0.514$.

\section{Supporting evidence}
\label{secquatro}
To validate the present construction, we compare our findings with relevant results in the literature, especially with experimental data.
\begin{figure*}
\includegraphics[width=\textwidth]{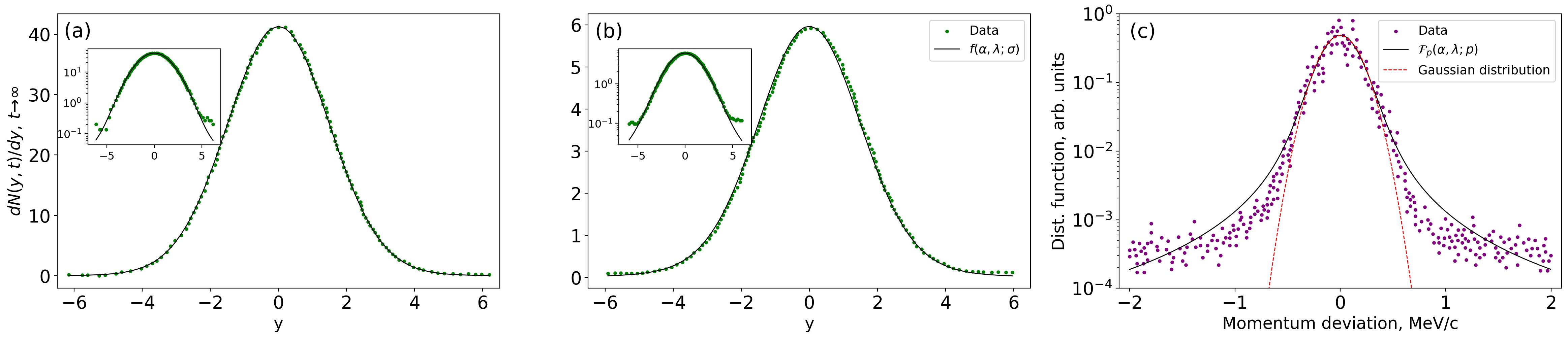}
\caption{\justifying 
Data fitting of theoretical and experimental data. (a), (b) Fits of the distribution data from heavy-ion collisions presented in Figs.~3 and 5 of Ref.~\cite{Wolschin2004Feb}, respectively, using Eq.~\eqref{eq:stablerapidity}~(black curves).
The parameter values are (for $Q$, see main text): $\alpha=1.959$, $\lambda=1.071$, $Q=156.792$ in (a) and $\alpha=1.86$, $\lambda=1.107$, $Q=23.37$ in (b).
The figures illustrate the probability distribution of 
rapidity $y$, which is obtained from $dN(y,t)/dy$ in the 
limit $t \to \infty$.
The insets highlight the strong agreement at the tails.
(c) Fitting of the experimental momentum deviation distribution data of antiproton cooling, as presented in Fig.~3 of Ref.~\cite{Nagaitsev2006Jan}, employing Eq.~\eqref{eq:rellsmomentum} (black curve), with $\alpha=1.736$, $\lambda=0.119$, $Q=0.206$.
The Gaussian distribution was considered in the original work~\cite{Nagaitsev2006Jan}. 
Remarkably, the tails observed experimentally are accurately described by the large-deviation regime of the relativistic stable distribution of momentum ${\mathcal F}_p(\alpha, \lambda; p)$.
}
\label{fig:expevidencesfitting}
\end{figure*}

First, a theoretical model based on rapidity as an independent random variable has been developed~\cite{Wolschin1999May,Wolschin2004Feb} to describe diffusion equilibration in heavy-ion collisions. The temporal evolution of the resulting distribution is illustrated in Fig.~2 of Ref.~\cite{Wolschin1999May} and Figs.~3 and 5 of Ref.~\cite{Wolschin2004Feb}.
Notably, the curves in these references are well fit by our $\alpha$-stable distribution of $\sigma$, 
multiplied by a numerical normalization factor $Q>0$.
This factor is necessary to adjust the probability distribution to the finite number of data points presented, 
as the standard normalization is for an infinite 
rapidity range.
A very good agreement with the plots in Ref.~\cite{Wolschin2004Feb} is shown in Fig.~\ref{fig:expevidencesfitting}(a) and (b).

Second, Ref.~\cite{Nagaitsev2006Jan} employed two 
different techniques \textemdash both relying on 
electron beams to generate a damping force \textemdash to
measure the distribution of momentum deviation 
during the cooling process of antiprotons.
One followed a two-step procedure, with the objective
of thermalizing antiprotons initially with high 
momentum deviations.
Consequently, this approach may suppress power-law trends typical of Lévy processes.
The other was implemented in a single step,
targeting antiprotons with small momentum deviations.
It resulted in a distribution exhibiting tail behavior,
as shown in Fig.~3 of Ref.~\cite{Nagaitsev2006Jan}. 
These tails were attributed to noise, speculated to
arise from the experimental setup. 
However, the overall behavior of the experimental curve in Fig.~3 of Ref.~\cite{Nagaitsev2006Jan} resembles the relativistic stable distribution of momentum shown in the inset of Fig.~\ref{fig:rel_energy_momentum}(b).
Indeed, the whole system is consistently driven out of
equilibrium via the injection of antiprotons, whose 
energy becomes dissipated through friction with the
electron beams. 
Such dynamics can lead to strong long-range correlations (in a relativistic context), and consequently, to heavy-tail behavior.
To re-examine and fit the data, we considered Eq.~\eqref{eq:rellsmomentum}, multiplied by a normalization
factor $Q>0$. 
As shown in Fig.~\ref{fig:expevidencesfitting}(c), Eq.~\eqref{eq:rellsmomentum} provides a significantly better fit to the empirical data than the Gaussian distribution in Ref.~\cite{Nagaitsev2006Jan}. 
This suggests that the observed tails might not be merely extraneous artifacts. 
Instead, they may indicate a stationary state described by a stable distribution of relativistic momentum.

We shall finally mention that similar reasoning related to the emergence of $\alpha$-stable distributions for stationary non-equilibrium states has been proposed in previous works for non-relativistic
phenomena~\cite{Barkai2004Jun,Pulvirenti2004Mar,Zanette2003Sep}.


\section{Conclusion}
We conclude this work by highlighting the technical and physical significance of our main finding, summarized in Eq.~(\ref{eq:rells}).
First, the approach introduced here provides a robust method for characterizing relativistic regimes in statistical systems.
In particular, the simple and easily calculated quantifiers $R(\alpha,\lambda)$ and $p_{r}$ are important for determining the correct values of the parameters $\alpha$ and $\lambda$ of the 
general $\mathcal{F}(\alpha,\lambda; \beta)$; and also for assessing the prevalence of relativistic effects in a given system.
Further, from Eqs.~\eqref{eq:rells}, \eqref{eq:rellsenergy}, and \eqref{eq:rellsmomentum}, we obtain a straightforward method to test empirical data for relativistic stable distributions.

Second, Jüttner and m-Jüttner distributions are often
criticized for their inability to explain particle 
emission from relativistic sources such as cosmic rays, 
quasars, and black holes, which are usually typified
by log-normal distributions \cite{Ioka2002Apr,Gaskell2004Aug,Gandhi2009May,Kunjaya2011Dec,Gladders2013May}. 
These rather important kinds of problems represent fundamental situations for testing relativistic stable distributions.
We emphasize that our approach has successfully described experimental results in particle physics. 
Thus, our framework offers new insights to understand relativistic statistical phenomena across domains, such as particle physics, plasma physics, and astrophysical processes.

Third, from a mathematical perspective, we remark that only in one dimension Eq.~\eqref{eq:rapaddrel} 
can be directly linked to Eq.~\eqref{eq:reladd}~\cite{fock1964,dunkel2007}. 
For dimensions $d > 1$, the relativistic velocity addition becomes significantly more complex: 
$\beta^2 = \frac{(\vec{\beta}_1 + \vec{\beta}_2)^2 -
(\vec{\beta}_1 \times 
\vec{\beta}_2)^2}{(1 + \vec{\beta}_1 
\cdot \vec{\beta}_2)^2}$. 
Furthermore, the stable distributions for $\alpha < 2$ 
become non-separable~\cite{uchaikin1999}. 
All this introduces additional complexity, requiring further studies for proper extension to $d > 1$, which is an ongoing investigation.


\begin{acknowledgements}
We are deeply grateful to G.~Kremer, A.~Mohammadi, E.~Barkai, P.~de Castro, and R.~Silva for enlightening discussions and comments.
We thank the Brazilian agency Conselho Nacional de Desenvolvimento Cient\'{\i}fico e Tecnol\'ogico (CNPq) for the grants 140921/2021-4 (LGBS), 307512/2023-1 (MGEL), 308840/2023-2 (EPR), 310928/2022-2 (EMFC), 302414/2022-3 (GMV), and 404577/2021-0 (Projeto Universal).
Also, LGBS acknowledges Coordenação de Aperfeiçoamento de Pessoal de Nível Superior (CAPES) for grant 88887.911602/2023-00, EPR acknowledges Funda\c{c}\~ao de Amparo \`a Ci\^encia e Tecnologia do Estado de Pernambuco (FACEPE), and EMFC acknowledges Funda\c{c}\~ao de Amparo \`a Ci\^encia e Tecnologia do Estado do Rio de Janeiro (FAPERJ) for partial support.
\end{acknowledgements}


\appendix


\section{\label{appen:invariant_functional}~Lorentz invariance of the distributions functional form}
Consider a (3+1)-Minkowski spacetime of coordinates
${x^\mu=(ct,x,y,z)}$ and metric ${\eta_{\mu\nu}=\mathrm{diag}(+,-,-,-)}$, where $\mu=0,1,2,3$.
Correspondingly, we define a phase space
$(x^\mu,p^\mu)$, where $p^\mu = (E/c,p_x,p_y,p_z)$ is the four-momentum.
Assume the inertial reference frames $S$ and $S'$, 
have the $x$ and $x'$-axes parallel to each other
and that the relative velocity $\mathbf{\beta}_0$ of $S'$ with respect to $S$ is along
such direction. Then their spacetime coordinates satisfy 
the Lorentz transformations $y' = y$, $z' = z$ and
\be \label{eq:Lorentzx}
x' = \gamma(\beta_0) \, (x - \beta_0 \, c t),
\ee
\be \label{eq:Lorentzt}
c t' = \gamma(\beta_0) \, (c t - \beta_0 \, x).
\ee
For a particle of momentum ${\bf p} = (p_x,p_y,p_z)$ 
and energy $E$ in $S$, in $S'$ these quantities 
are $p'_y = p_y$, $p'_z = p_z$ and
\be \label{eq:Lorentzpx}
p'_x = \gamma(\beta_0) \, (p_x - \beta_0  \, E/c),
\ee
\be \label{eq:LorentzE}
E'/c = \gamma(\beta_0) \, (E/c - \beta_0 \, p_x).
\ee
The above relations can be given solely in terms of 
momentum since ${p_0=\gamma(\beta_0) \, \beta_0 \, m
\, c}$ and ${E(p) = \gamma(p) \, m \, c^2}$.

To construct a Lorentz‐invariant phase‐space volume element, we define
\be \label{eq:invariant_ph_space_volume}
d\Gamma := 2 \,
\delta(E^2 - p^2 c^2 - m^2 c^4) \,
\Theta(E) \, d^4{\rm x} \, d^4{\rm  p},
\ee
where 
$
{d^4{\rm x} = d(ct) \, d^3x}
$
and
$
{d^4{\rm  p} = d(E/c) \, d^3p}
$
are the spacetime and energy-momentum volume elements. 
Here, $\delta(\cdot)$ is the Dirac delta distribution and 
$\Theta(\cdot)$ is the Heaviside function. 
If a particle has rest mass, the on-shell condition $E^2 = p^2 c^2 - m^2 c^4$ holds.
We impose it together with energy positivity through the
$2 \, \delta(E^2 - p^2 c^2 - m^2 c^4)\,\Theta(E)$.
The argument of the delta distribution is itself a Lorentz scalar, and the energy positivity condition is not affected by Lorentz transformations.
Additionally, both volume elements $d^4{\rm x}$ and $d^4{\rm p}$ are Lorentz-invariant since the Jacobian under Lorentz transformations is $1$.
Hence, we conclude that Eq.~\eqref{eq:invariant_ph_space_volume} is a Lorentz-invariant phase space volume element.

Now, suppose an arbitrary probability distribution $f(x^\mu,p^\mu)$ is constructed on the SR phase space.
Therefore, the proper measure of $f$ reads 
$$
f(x^\mu,p^\mu) \, d\Gamma.
$$
Consider the identity ${ \delta(H(z)) = \sum_i \frac{\delta(z-z_i)}{|H'(z_i)|} }$, where $H(z)$ is a differentiable function and ${H(z_i)=0}$.
From this, we find
$$
{2 \, \delta(E^2 - p^2 c^2 - m^2 c^4) \, \Theta(E)} = \frac{\delta(E - \sqrt{p^2c^2 + m^2 c^4})}{E(p)},
$$
with ${E(p)=\sqrt{p^2c^2 + m^2 c^4}=\gamma(p) \, m \, c^2}$.
Therefore, we can write the associated total probability as
\be \label{eq-measure}
\begin{aligned}
\int f(x^\mu,p^\mu) \, d\Gamma & =\int f(x^\mu,p^\mu)\mid_{E=E(p)} \, 
\frac{d^4{\rm x} \, d^3p}{E(p)} \\
&= \int \frac{f(x^\mu,{\bf p})}{E(p)} 
\, d^4{\rm x} \, d^3p.
\end{aligned}
\ee
with ${f(x^\mu,{\bf p}) = f(x^\mu,p^\mu)\mid_{E=E(p)}}$. 
To the probability in Eq.~\eqref{eq-measure} to be a Lorentz scalar, $f(x^\mu,{\bf p})$ needs to have at least its functional form preserved under Lorentz transformations.
Since the integration runs over the entire phase space, shifting the distribution of independent variables by a constant does not affect the integral (see discussion below).

Given that the addressed distributions are one-dimensional in momentum, we use the second expression in 
Eq.~\eqref{eq-measure} with volume elements $d^2{\rm x}=d(ct) \, dx$ and $d^2{\rm p}=d(E/c) \, dp$, so that
\be \label{eq:invariantN_1_1}
\int f(x^\mu,p^\mu) \, d\Gamma 
= \int \frac{f(x^\mu,p)}{E(p)} 
\, d^2{\rm x} \, dp.
\ee
We should emphasize that since the distributions 
considered here, Eq.~\eqref{eq:rellsmomentum} ---
 as well as \Juttner distribution --- are constructed 
over the momentum space, they implicitly assume homogeneity in
space and time.
Hence, the spacetime volume element can be integrated out 
of the expression, without affecting its Lorentz-scalar property.

In particular, the integration of the energy takes the volume element ${2 \, \delta(E^2 - p^2 c^2 - m^2 c^4) \, \Theta(E)} \, d^2{\rm  p}$ to the well-known Lorentz invariant quantity $\frac{dp}{E(p)}$~\cite{Debbasch2001Dec,Debbasch2009Apr,Debbasch2009May}.
In one dimension, this is also obtained from the definition of rapidity $\sigma(p) = \sinh^{-1}(p/mc)$, so that ${d \sigma = \frac{dp}{E(p)/c} }$.
In addition, the equivalent quantity in the relativistic velocity space is the invariant volume element $\gamma^2(\beta) \, d\beta$~\cite{fock1964}, arising from $p=\gamma(\beta) \, \beta \, m \, c$.
This volume element is present in Eq.~\eqref{eq:rells}.

From the discussion above, we can verify whether \Juttner-like and relativistic stable distributions are consistent with a Lorentz-invariant phase space volume.
This should lead to a Lorentz-scalar total probability.

\subsection{\label{appen:invariant_functional_Juttner}
\Juttner-like distributions}
The one-dimensional generalized \Juttner distribution
in phase space is~\cite{dunkel2007}
\be \label{eq:genJuttnerphspace}
f_{\eta}(p) \, dp = \frac{1}{Z_\eta \,
E^{\eta}(p)}
\, \exp \left( -\frac{E(p)}{k_{\rm B} T} \right) \, dp,
\ee
where $Z_\eta$ is a normalization factor calculated for each $\eta$.
It reduces to the classic \Juttner distribution for ${\eta=0}$ and to the m-\Juttner for ${\eta=1}$.
For Eq. (\ref{eq:genJuttnerphspace}), it is adopted Landsberg's perspective, namely, that temperature is a Lorentz scalar~\cite{landsberg1967}.

In SR, the energy transforms by adding a momentum linear dependence, as shown in Eq.~\eqref{eq:LorentzE}.
Consequently, because the way $E(p)$ enters into  Eq.~\eqref{eq:genJuttnerphspace}, the functional form
of $f_{\eta}(p) \, dp$
changes under Lorentz transformations; presumably with the associated total probability depending on the relative velocity.
Thus, the total probability obtained from it is not a Lorentz scalar.

\subsection{\label{appen:invariant_functional_RLS}Relativistic stable distributions}
Consider the relativistic stable distribution in the SR phase space in Eq.~\eqref{eq:rellsmomentum},
$$
{\cal F} (\alpha,\lambda;p) \, dp = \frac{f(\alpha,\lambda;\sigma(p))}{E(p)/c} \, dp.
$$
Then, we can write
\be
\begin{aligned}
\label{eq:relstabdistinvariance}
\int {\cal F}(\alpha,\lambda;p') \, dp' 
 &= \int \frac{f(\alpha,\lambda;\sigma(p'))}
{E'(p')/c} \, dp' \\  
&= \int \frac{f(\alpha,\lambda;\sigma(p)-\sigma(p_0))}{E(p)/c} \, dp  
\nonumber \\
&= \int f(\alpha,\lambda;\sigma-\sigma_0) 
\, d\sigma \\
&= \int f(\alpha,\lambda;{\bar \sigma})
 \, d{\bar \sigma} \\
&= \int \frac{f(\alpha,\lambda;\sigma(p))}{E(p)/c} \, dp 
\nonumber \\
&= \int {\cal F}(\alpha,\lambda;p) \, dp.
\end{aligned}
\ee
Above, we have (i) applied the Lorentz transformations; 
(ii) switched from momentum to rapidity space;
(iii) performed the variable change $\bar{\sigma} = \sigma-\sigma_0$;
(iv) and finally went back from the rapidity to the momentum space.
We remark that since there is no physical constraint in the phase space, step (iii) does not change the integration limits.
Thus, the total probability $\int {\cal F}(\alpha,\lambda;p) \, dp$  is a Lorentz scalar because the functional form of $f(\alpha,\lambda;\sigma)$ is preserved under Lorentz transformations.

Notwithstanding, we should emphasize that although the present results are an important contribution to the general discussion of statistical measures and distributions in SR, this subject cannot be considered settled.
Stable distributions universally characterize the stationary state of a system~\cite{Barkai2003Nov,Barkai2004Jun}.
Only for the specific case $\alpha=2$ the distribution becomes Gaussian and thus describes the system in equilibrium, yielding the Maxwell–Boltzmann distribution.
The relation of phase space with distributions of stationary states is still open to further exploration.


\section{\label{appen:reladdN}~The relativistic addition of $N$ velocities}
From the Lorentz transformations in Eqs.~\eqref{eq:Lorentzx} and \eqref{eq:Lorentzt}, one easily obtains the relativistic velocity addition law
\be\label{eq:reladdappend}
\beta = \beta_1 \oplus \beta_2' = 
\frac{\beta_1 + \beta_2'}{1 + \beta_1 \beta_2'}.
\ee
This is equivalently written as
$$
\frac{1+\beta}{1-\beta} = \left(\frac{1+\beta_1}
{1-\beta_1}\right)\left(\frac{1+\beta_2'}{1-\beta_2'}\right).
$$
Now, if one defines 
$\beta_2' = \frac{\beta_2 + \beta_3'}{1 + \beta_2\,\beta_3'}$, 
then Eq.~\eqref{eq:reladdappend} becomes
$$
\beta =
\frac{\beta_1 + \beta_2 + \beta_3' + \beta_1 \beta_2 \beta_3'}
{1 + \beta_1 \beta_2 + \beta_1 \beta_3' + \beta_2 \beta_3'},
$$
which can be written as
$$
\frac{1+\beta}{1-\beta} = \left(\frac{1+\beta_1}{1-\beta_1}\right) \left(\frac{1+\beta_2}{1-\beta_2}\right) 
\left(\frac{1+\beta_3'}{1-\beta_3'}\right).
$$
By iterating this procedure
(with $\beta_j'$ defined recursively by $\beta_j' = \beta_j \oplus \beta_{j+1}'$ for $j=2,\dots,N-1$, and $\beta_N' = \beta_N$), we
conclude that 
\be
\beta = 
\displaystyle \oplus_{j=1}^{N} \beta_j 
= \frac{ \displaystyle \sum_{j \text{ odd}}^N  \, \,
\displaystyle \sum_{n_1 < n_{2} < \cdots < n_j}^{N} 
\beta_{n_1} \dots \beta_{n_j}}{1 +
\displaystyle \sum_{j \text{ even}}^N \, \, 
\sum_{n_1 < n_{2} < \cdots < n_j}^{N}
\beta_{n_1} \dots \beta_{n_j}}.
\ee


\section{\label{appen:stabcondproof}~Proof of the relativistic stability condition}
The relativistic stability condition proposed in the main text can be shown to hold by establishing the existence of an isomorphism between the group of rapidities and the group of relativistic velocities.

Let ${\mathcal G}_\beta = (V, \oplus)$ and ${\mathcal G}_\sigma = (S, +)$ be the groups of relativistic velocities and rapidities, where $V = \{\beta \in (-1,1)\}$ and $S = \{\sigma \in (-\infty,\infty)\}$.
Their respective binary operations $\oplus$ and $+$ correspond to the relations in Eqs.~\eqref{eq:reladd} and \eqref{eq:rapaddrel}.
Consider the map $\Sigma\colon V \to S$ defined by Eq.~\eqref{eq:rapidity}. 
We shall demonstrate that $\Sigma$ is an isomorphism. 

Closure under $\oplus$ follows directly from Eq.~\eqref{eq:rapaddrel}.
Moreover, the identity element of $\mathcal{G}_\sigma$ is $0\in S$, and that of $\mathcal{G}_\beta$ is $0\in V$.
Hence, through $\Sigma$, the identity element of $\mathcal{G}_\beta$ is mapped into the identity element of $\mathcal{G}_\sigma$.
For each $\beta\in V$ with inverse $-\beta$, the image $\Sigma(\beta)$ has inverse $-\Sigma(\beta)$, showing that $\Sigma$ preserves inverses.
Lastly, one directly verifies that the function $\Sigma(\beta) = \tanh^{-1}(\beta)$ is bijective.
Therefore, $\Sigma$ is an isomorphism.

\begin{figure}[ht!]
\includegraphics[width=0.47\textwidth]{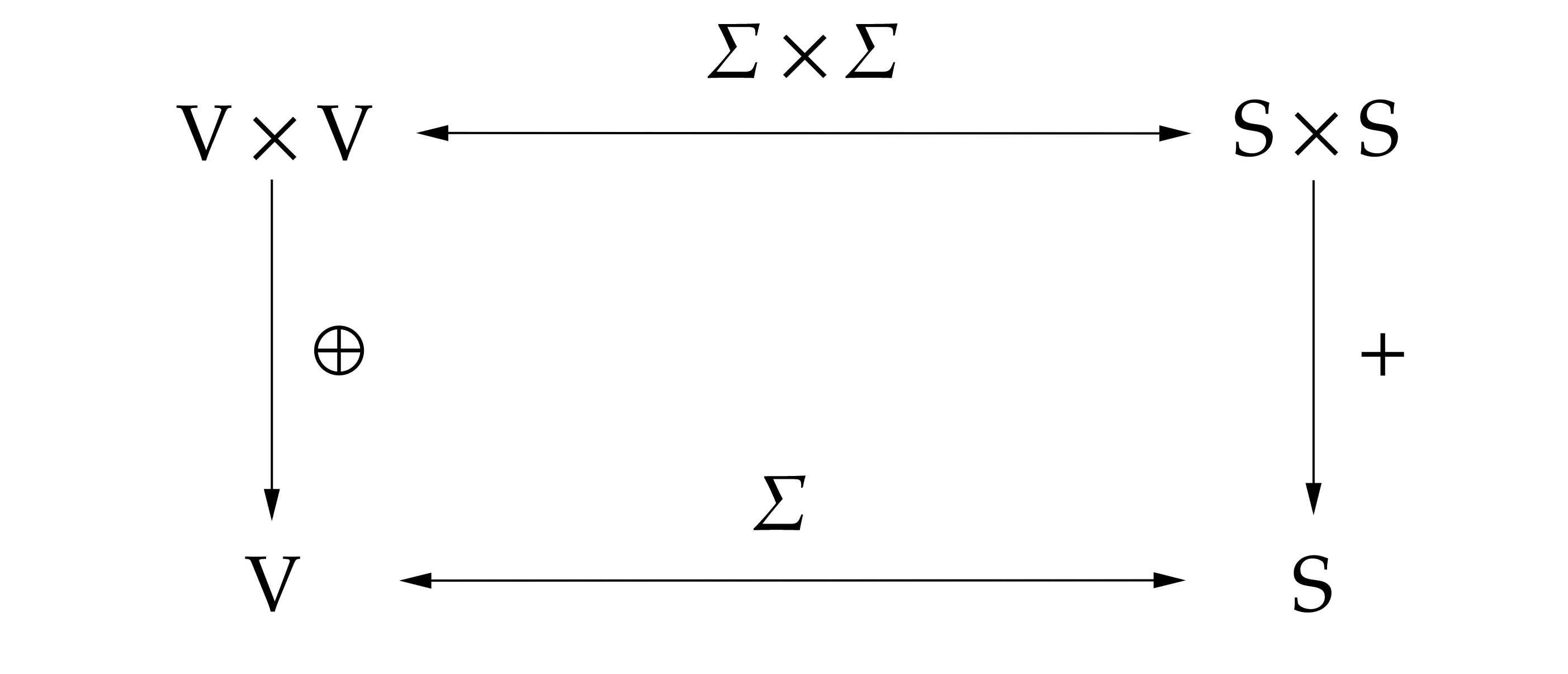}
\caption{\justifying
Diagrammatic representation of the isomorphism $\Sigma$ between the group of relativistic velocities $\mathcal{G}_\beta$ and the group of rapidities $\mathcal{G}_\sigma$. 
Since there exists the inverse $\Sigma^{-1}$, the distribution $\mathcal{F}(\alpha,\lambda;\beta)$ can be constructed either over the group $\mathcal{G}_\sigma$ or the group $\mathcal{G}_\beta$.
}
\label{fig:lilsdiagram}
\end{figure}

Since $\Sigma$ is an isomorphism, there exists an inverse $\Sigma^{-1}$. 
Thus, we can derive $\mathcal{F}(\alpha, \lambda; \beta)$ either from the relativistic stability condition of $\beta$ or from the stability condition of $\sigma$. 
A schematic of the proof is shown in Fig.~\ref{fig:lilsdiagram}.


\section{\label{appen:meanv2}~Deriving the velocity second moment for the $\alpha=1$ case}
Consider the relativistic $\alpha$‑stable velocity distribution at $\alpha=1$.
By definition, its variance is given by
\be \label{eq:1meanv2}
\begin{aligned}
\mean{ \beta^2 }(\lambda) & = \int_{-1}^{1} \beta^2 \, \gamma^2(\beta) \, \left(\frac{ \lambda / \pi} 
{ \sigma(\beta)^2 + \lambda^2 }\right) \, d\beta \\
 &  = \frac{ 1 }{ \pi } \int_{-\infty}^{\infty} \frac{
 \tanh^2(\lambda \, w)} { w^2 + 1 } \, d w.
\end{aligned}
\ee
The second integral follows from the change of variables
$w = \sigma(\beta) / \lambda = \tanh^{-1}(\beta) /
 \lambda$.

Noting that $\frac{d}{dx}\tanh(a x)=a\,(1-\tanh^2(a x))$, we rewrite
\be 
\label{eq:2meanv2}
\mean{ \beta^2 }(\lambda)
= 1 - \frac{2}{\lambda \, \pi} 
\int_{-\infty}^{\infty}
 \frac{ w \tanh(\lambda \, w) } { (w^2 + 1)^2 } \, d w.
\ee
The hyperbolic tangent function admits the series representation
\be \label{eq:tanhgenhyper}
\begin{aligned}
\tanh(z) &= \frac{8z}{4 z^2 + \pi^2} \,
{}_3 F_2 \left(^{1,\frac{1}{2}-\frac{iz}{\pi},\frac{1}{2}+\frac{iz}{\pi}}_{\frac{3}{2}-\frac{iz}{\pi},\frac{3}{2}+\frac{iz}{\pi}}\mid 1 \right) \\
& = \frac{8z}{4 z^2 + \pi^2} \sum_{n=0}^{\infty} \frac{\pi^2+4z^2}{(1+2n)^2\pi^2+4z^2} \, ,
\end{aligned}
\ee
where ${}_3F_2$ denotes the generalized hypergeometric function ${}_p F_q\left(^{\vec{a}}_{\vec{b}} \mid z \right)$, with $p=3$ and $q=2$~\cite{erdelyi1953}.

Substituting the series from Eq.~\eqref{eq:tanhgenhyper} into Eq.~\eqref{eq:2meanv2} and integrating term by term yields
\be
\mean{ \beta^2 } = 1 -\frac{2}{\pi^2} \sum_{n=0}^\infty \frac{1}{(n+\lambda/\pi+1/2)^2}.
\ee
The resulting series is the first‑order polygamma function, $\psi^{(1)}(z) = \frac{d^2}{dz^2} \ln(\Gamma(z)) = \sum_{n=0}^\infty \frac{1}{(z+n)^2}$~\cite{abramowitz1968}. 
Therefore,
$$
\mean{ \beta^2 }(\lambda) = 1 - \frac{2}{\pi^2} \psi^{(1)} \left(\frac{\lambda}{\pi} 
+ \frac{1}{2} \right).
$$
In the limit $\lambda\to\infty$, one finds $\mean{\beta^2} \to 1$, confirming the variance remains finite.


\smallskip
\begin{widetext}
\section{\label{appen:alpharational}~Analytical formulas for certain rational $\alpha$'s}
Closed expressions exist for $\alpha=2/M$, where $M=1,2,3,\dots$; the cases $M=1$ and $M=2$ were discussed above.
The derivation of the traditional distribution $f$ can be found in \cite{Crisanto-Neto2016}, where the distributions are written as a sum of generalized hypergeometric functions. 
Following \cite{Crisanto-Neto2016}, the relativistic distribution ${\mathcal F}$ is given by
\be\label{eq:lilsalpharational}
\mathcal{F}(2/M,\lambda;\beta) = 
\frac{M\Gamma\left(\frac{M}{2}\right)}{2\pi\lambda C}
\gamma^2(\beta) \sum_{j=1}^{M-1}
\Gamma\left(b_j-\frac{1}{2}\right)
\frac{C_j/M^{\frac{M}{2}+j}}{\left(\sigma^2(\beta)/
4\lambda^2\right)^{\frac{1}{2}+\frac{j}{M}}} {}_1 F_{M-2}
\left(\begin{array}{c}
b_j - \frac{1}{2} \\
\vec{a}^{(j)}
\end{array} 
\mid \frac{(-1)^{M-1}4\lambda^2}{M^M\sigma^2(\beta)}\right),
\ee
where 
$$
{C= \displaystyle\prod_{j=1}^{M-1} \Gamma\left(b_j-\frac{1}{2}\right)},
\qquad
{C_j=\displaystyle\prod_{i=1, i \neq j}^{M-1}
\Gamma\left(b_i-b_j\right)},
\qquad
{b_j=1+\frac{j}{M}}$$
and 
$
\vec{a}^{(j)}= \left(b_j-\frac{1}{M}, b_j-\frac{2}{M}, \ldots, b_j-\frac{(j-1)}{M}, b_j-\frac{(j+1)}{M}, b_j-\frac{(j+2)}{M}, \ldots, b_j-\frac{(M-1)}{M}\right).
$

Below, we present the relativistic $\alpha$-stable 
distributions for various rational values of $\alpha$. 
We also show their graphs in Fig.~\ref{fig:lilsrationalalphas}.
The traditional $\alpha$‑stable distributions for these values appear, for example, in Ref.~\cite{lee2010}.
\begin{itemize}
\item $\alpha=1/3$
\be
\mathcal{F}( 1/3, \, \lambda; \, \beta) = \operatorname{Re}\left[\frac{2 \lambda^{1/2} e^{-i \pi / 4} \gamma^2(\beta) }{3 \sqrt{3} \pi|\sigma(\beta)|^{3 / 2} } S_{0,1 / 3}\left(\frac{2 \lambda^{1/2} e^{i \pi / 4} }{3 \sqrt{3} |\sigma(\beta)|^{1 / 2} }\right)\right].
\ee


\item $\alpha=1/2$
\be
\begin{aligned}
\mathcal{F}( 1/2, \, \lambda; \, \beta) = & \frac{ \lambda^{1/2} \gamma^2(\beta) |\sigma(\beta)|^{-3 / 2} }{ \sqrt{2 \pi}} \sin \left( \frac{\lambda }{4|\sigma(\beta)|} \right) \left(\frac{1}{2} - \mathrm{S}\left( \sqrt{\frac{\lambda }{2 \pi|\sigma(\beta)|}} \right) \right) \\
 & + \frac{ \lambda^{1/2} \gamma^2(\beta) |\sigma(\beta)|^{-3 / 2} }{ \sqrt{2 \pi} } \cos \left( \frac{ \lambda }{4|\sigma(\beta)| } \right) \left( \frac{1}{2 } - \mathrm{C} \left(
\sqrt{\frac{\lambda }{2 \pi|\sigma(\beta)|}} \right) \right),
\end{aligned}
\ee
where 
$\mathrm{C}(z)=\int_{0}^{z}\cos\bigl(\tfrac{\pi t^2}{2}\bigr)\,dt$ 
and 
$\mathrm{S}(z)=\int_{0}^{z}\sin\bigl(\tfrac{\pi t^2}{2}\bigr)\,dt$.


\item $\alpha=2/3$
\be
\mathcal{F}(2/3, \, \lambda; \, \beta) = \frac{ \gamma^2(\beta) |\sigma(\beta)|^{-1}}{2 \sqrt{3 \pi }}  \exp\left( \frac{ 2 \lambda^2 / 27 }{ \sigma^2(\beta) } \right) \mathrm{W}_{ - \frac{1}{2} , \frac{1}{6} }\left(\frac{ 4 \lambda^2 / 27 }{ \sigma^2(\beta)} \right),
\ee
where $\mathrm{W}_{a,b}(z)$ is the Whittaker function.


\item $\alpha=4/3$
\be
\begin{aligned}
\mathcal{F}( 4/3, \, \lambda; \, \beta) & = \frac{3^{5 / 4} \gamma^2(\beta) \Gamma( \frac{7}{12} ) \Gamma( \frac{11}{12} ) }{ 2^{5 / 2}\sqrt{\pi \lambda^2} \Gamma( \frac{6}{12} ) \Gamma( \frac{8}{12} )} {}_2F_2 \left( \begin{array}{c}
 \frac{7}{12}, \frac{11}{12} \\
 \frac{6}{12}, \frac{8}{12} 
\end{array}  \mid \frac{3^3 \sigma^4(\beta)}{2^8} \right) \\
 & -\frac{3^{11 / 4} \gamma^2(\beta) |\sigma(\beta)|^3}{2^{13 / 2} \sqrt{\pi} \lambda^4 } \frac{\Gamma(\tfrac{13}{12})\,\Gamma(\tfrac{17}{12})}{\Gamma(\tfrac{18}{12})\,\Gamma(\tfrac{15}{12})} {}_2F_2 \left(
\begin{array}{c}
 \frac{13}{12}, \frac{17}{12} \\
 \frac{18}{12}, \frac{15}{12} 
\end{array}  \mid \frac{3^3 \sigma^4(\beta)}{2^8}\right).
\end{aligned}
\ee


\item $\alpha=3/2$, corresponding then to the relativistic Holtsmark distribution
\be
\begin{aligned}
\mathcal{F}(3/2, \lambda;\beta) = & \frac{\Gamma(\frac{5}{3}) \gamma^2(\beta) }{\pi \lambda}  {}_{2} F_{3}\left( 
\begin{array}{c}
 \frac{5}{12}, \frac{11}{12} \\
 \frac{1}{3}, \frac{1}{2}, \frac{5}{6} 
\end{array}  \mid -\frac{2^{2} {\sigma (\beta)}^{6}}{3^{6} \lambda^6 }\right) 
 -\frac{ \gamma^2(\beta) {\sigma (\beta)}^{2} }{3 \pi \lambda^3 } { }_{3} F_{4}\left(\begin{array}{c}
 \frac{3}{4}, 1, \frac{5}{4} \\
 \frac{2}{3}, \frac{5}{6}, \frac{7}{6}, \frac{4}{3} 
\end{array}  \mid -\frac{2^{2} {\sigma (\beta)}^{6}}{3^{6} \lambda^6}\right) \\
 & +\frac{7 \, \Gamma(\frac{4}{3}) \gamma^2(\beta) {\sigma (\beta)}^{4} }{ 3^{4} \pi \lambda^5} {}_{2} F_{3}\left(\begin{array}{c}
 \frac{13}{12}, \frac{19}{12} \\
 \frac{7}{6}, \frac{3}{2}, \frac{5}{3} 
\end{array}  \mid -\frac{2^{2} {\sigma(\beta)}^{6}}{3^{6} \lambda^6}\right).
\end{aligned}
\ee
\end{itemize}
\begin{figure*}
\includegraphics[width=0.47\textwidth]{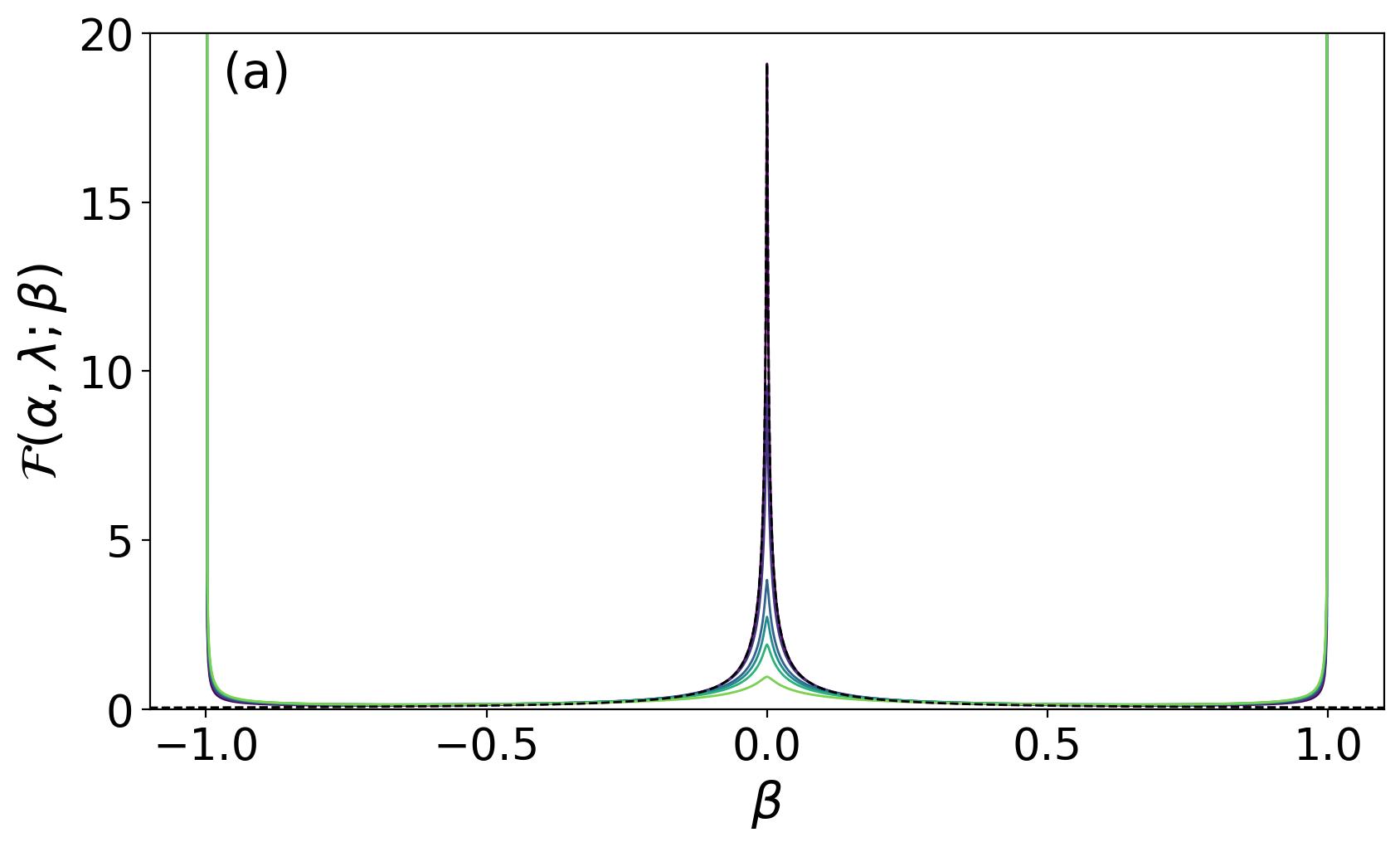}
\includegraphics[width=0.47\textwidth]{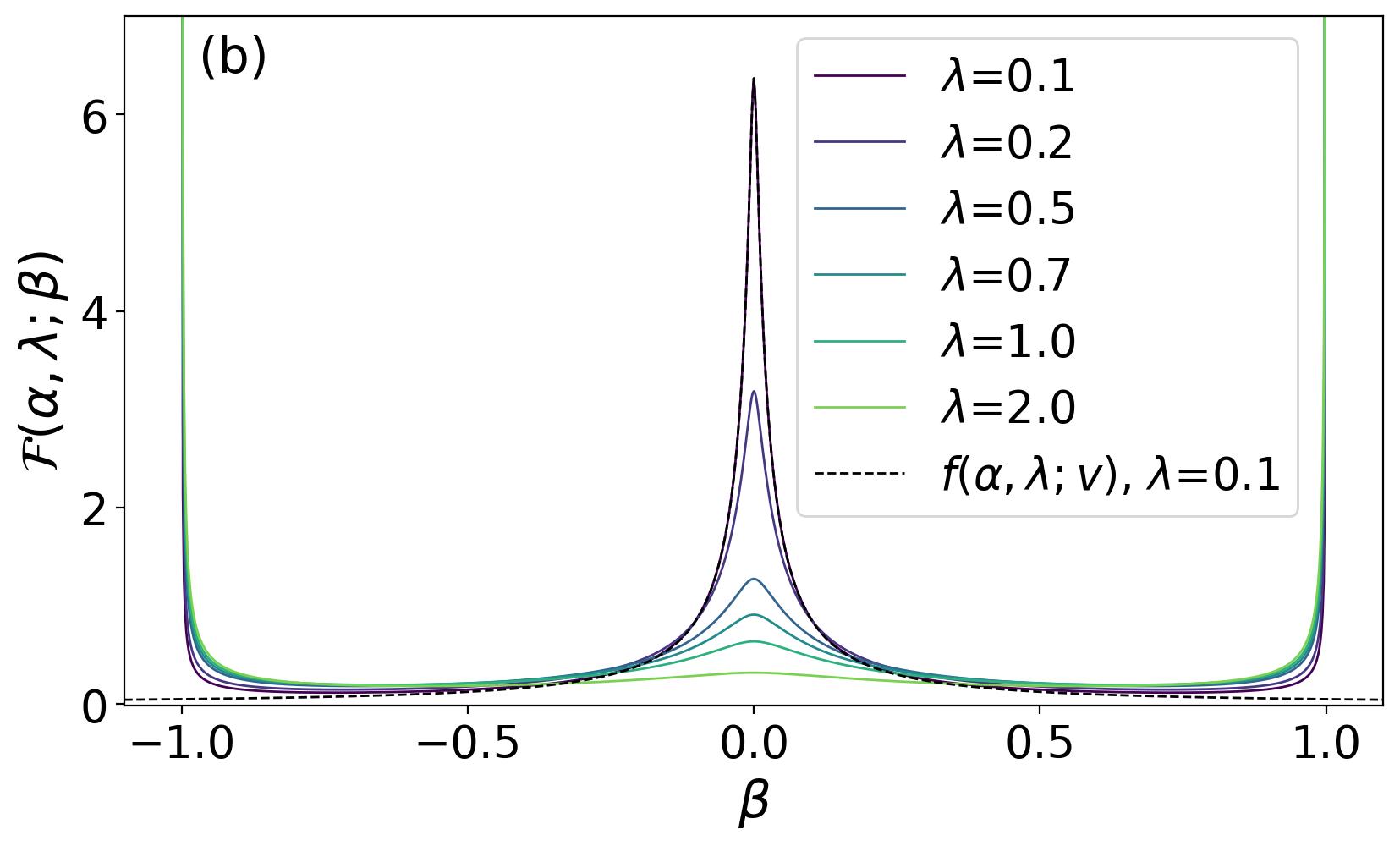}
\includegraphics[width=0.47\textwidth]{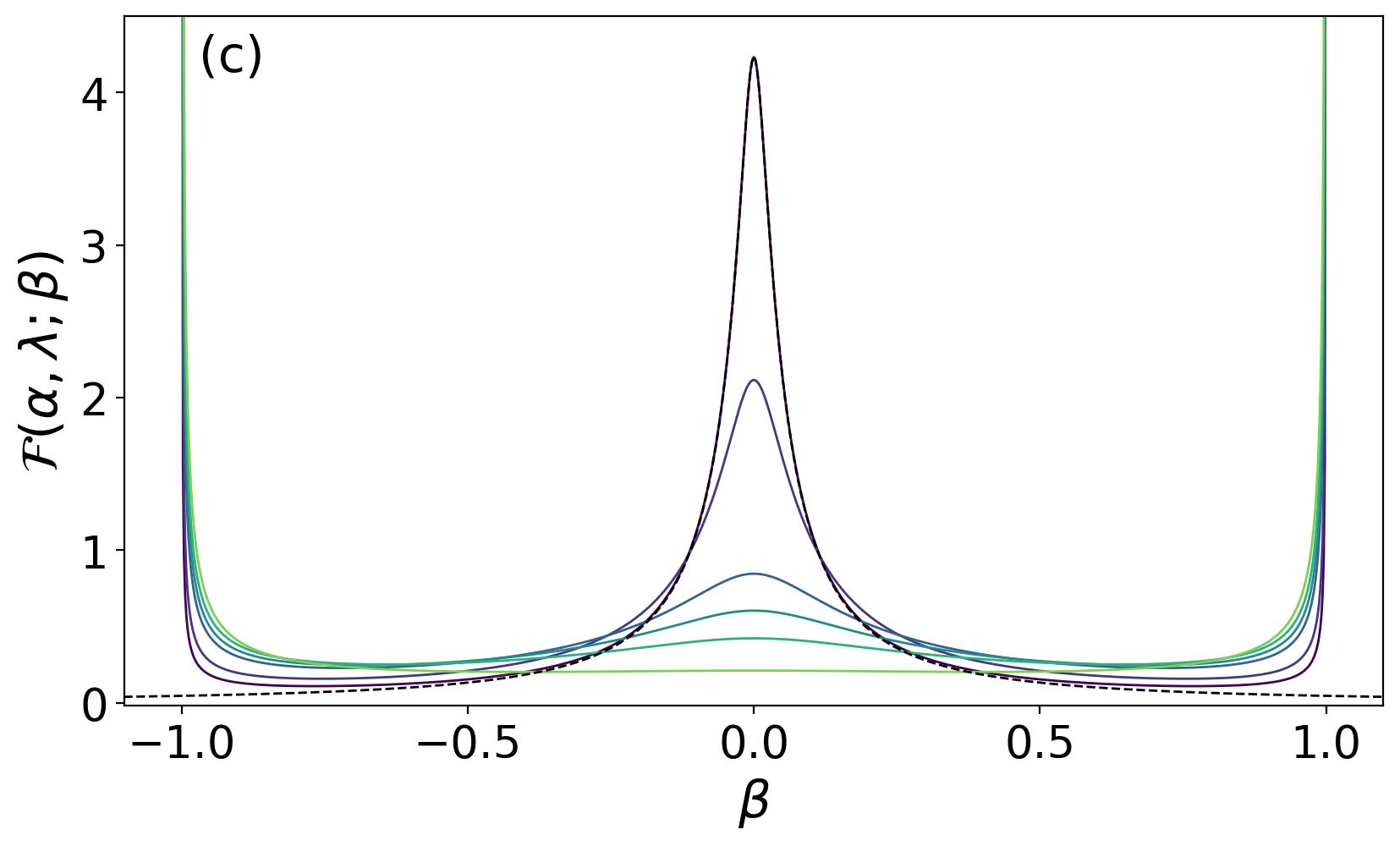}
\includegraphics[width=0.47\textwidth]{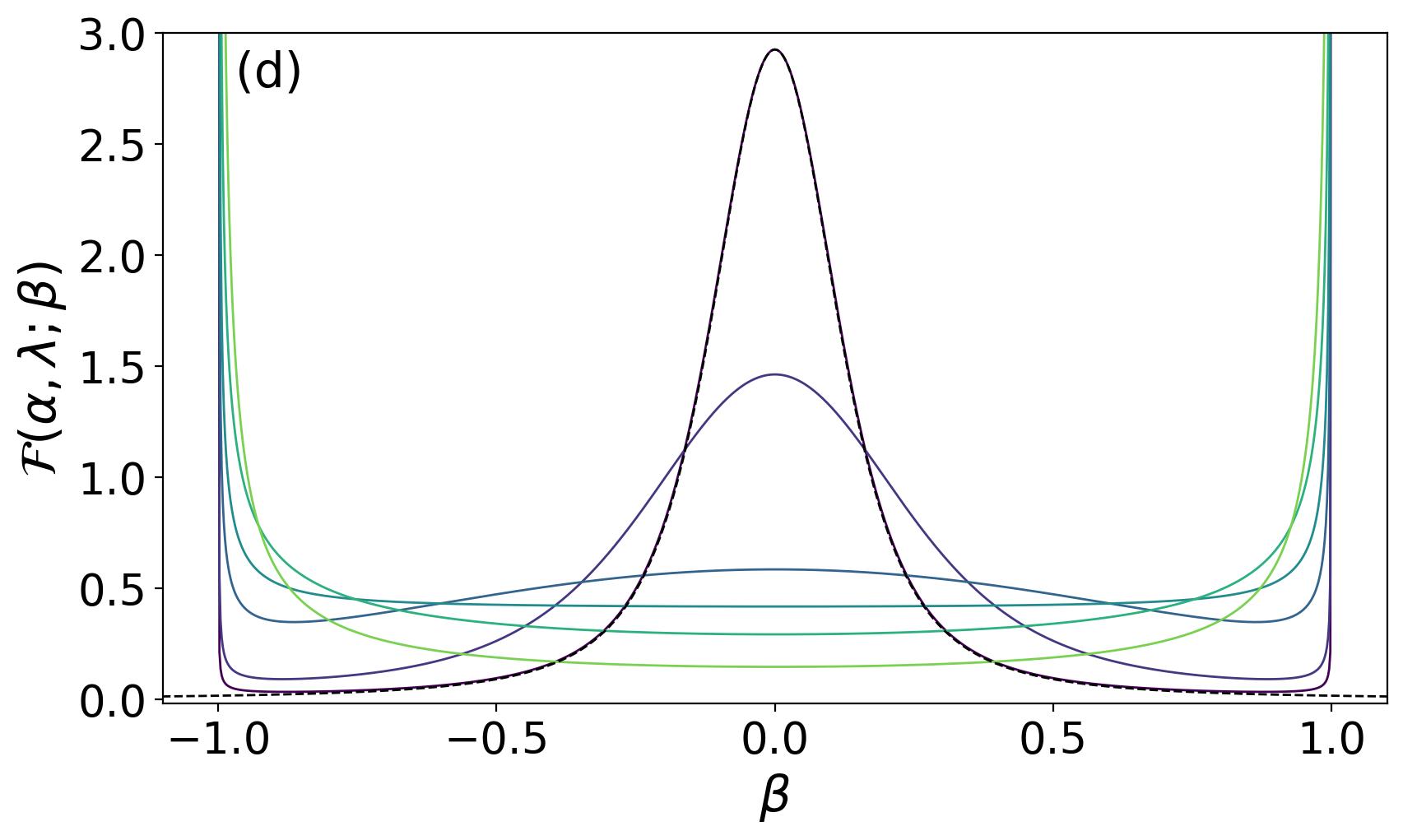}
\includegraphics[width=0.47\textwidth]{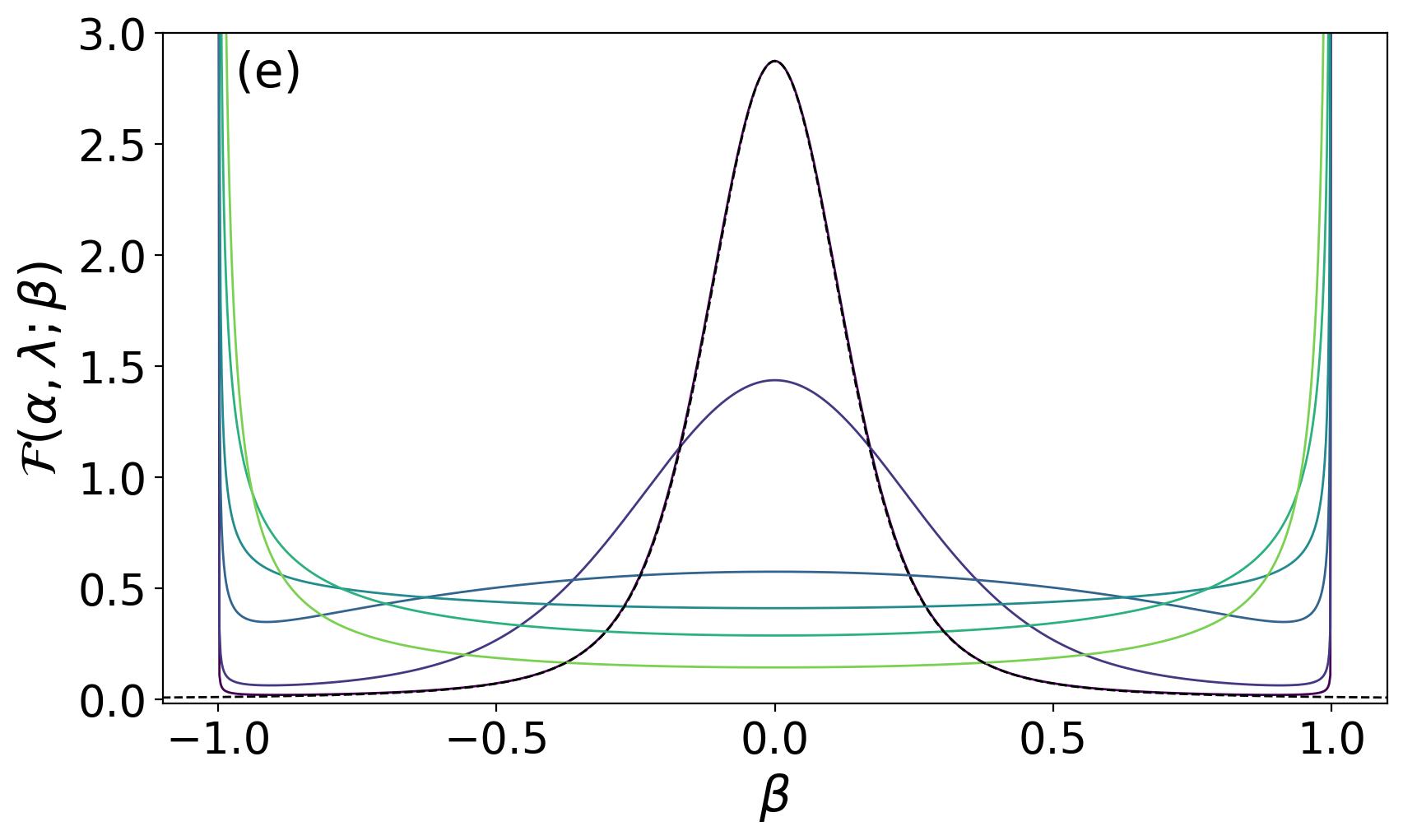}
\caption{\justifying
Relativistic stable distributions for (a)-(e) $\alpha = 1/3,\,1/2,\,2/3,\,4/3,\,3/2$ (solid curves) and their corresponding non-relativistic stable distributions (dashed curves). 
For all $\alpha>2/3$, the distribution transitions from trimodal to bimodal behavior with $\lambda$. 
The bimodal behavior is absent for $\alpha\leq2/3$, as then $\lambda$ remains below the critical value $\lambda_c^{(2/3)}=2.56$.
}
\label{fig:lilsrationalalphas}
\end{figure*}
\end{widetext}


\clearpage
\bibliography{Rel_LS_distributions}

\end{document}